\documentclass{aa}  

\usepackage{graphicx}
\usepackage{xcolor}
\usepackage{array,booktabs}
\usepackage{txfonts}
\usepackage{hyperref}
\hypersetup{colorlinks=true, citecolor=blue}

\usepackage{etoolbox}

\def\cigale {\textsc{cigale}}
\def\cloudy {\textsc{cloudy}}
\def\chisquare {$\chi^2$}
\def\cmt {cm$^{-3}$}
\def\lmfit {\textsc{lmfit}}
\def\kms {km\,s$^{-1}$}
\def\myr {M$_\odot$\,yr$^{-1}$}
\def\myrkpc {M$_\odot$\,yr$^{-1}$\,kpc$^{-2}$}
\def\magphys {\textsc{magphys}}
\def\mappings {\textsc{mappings v}}
\def\mappingsiii {\textsc{mappings iii}}
\def\pycasso {\textsc{pycasso}}
\def\pyneb {\textsc{pyneb}}
\def\scipy {\textsc{scipy}}
\def\starlight {\textsc{starlight}}

\newcommand{\nii}{[N\,{\sc ii}]}
\newcommand{\oi}{[O\,{\sc i}]}
\newcommand{\oiii}{[O\,{\sc iii}]}
\newcommand{\sii}{[S\,{\sc ii}]}

\begin{document}

   \title{VALES VIII: Weak ionized gas outflows in star-forming \\ galaxies at $z \sim 0.15$ traced with VLT/MUSE}


   \author{Guilherme S. Couto\inst{1} \and
	   Thomas M. Hughes\inst{2,3,4,5} \and
	   Médéric Boquien\inst{1} \and \\
	   Eduardo Ibar\inst{3} \and
	   Sébastien Viaene\inst{6} \and 
	   Roger Leiton\inst{7} \and
	   Yongquan Xue\inst{4,5}}

   \institute{Centro de Astronomía (CITEVA), Universidad de Antofagasta, Avenida Angamos 601, Antofagasta, Chile\\
   \email{guilherme.couto@uantof.cl}
   \and
   Chinese Academy of Sciences South America Center for Astronomy, China-Chile Joint Center for Astronomy, Camino El Observatorio 1515, Las Condes, Santiago, Chile \\
   \and
   Instituto de Física y Astronomía, Universidad de Valparaíso, Avda. Gran Bretaña 1111, Valparaíso, Chile\\
   \and
   CAS Key Laboratory for Research in Galaxies and Cosmology, Department of Astronomy, University of Science and Technology of China, Hefei 230026, China\\
   \and
   School of Astronomy and Space Science, University of Science and Technology of China, Hefei 230026, China\\
   \and
   Boltzmann BV, Belgium\thanks{The opinions expressed in this paper are solely of the author and do not represent the views of the company.}\\
   \and
   CePIA, Departamento de Astronomía, Universidad de Concepción, Casilla 160-C, Concepción, Chile\\
   }

   \date{Received ???; accepted ???}

 
  \abstract{

  We characterize the ionized gas outflows in 15 low-redshift star-forming galaxies, a Valparaíso ALMA Line Emission Survey (VALES) subsample, using MUSE integral field spectroscopy and GAMA photometric broadband data. We measure the emission-line spectra by fitting a double-component profile, with the second and broader component being related to the outflowing gas. This interpretation is in agreement with the correlation between the observed star-formation rate surface density ($\Sigma_{\mathrm{SFR}}$) and the second-component velocity dispersion ($\sigma_{\mathrm{2nd}}$), expected when tracing the feedback component. By modelling the broadband spectra with spectra energy distribution (SED) fitting and obtaining the star-formation histories of the sample, we observe a small decrease in SFR between 100 and 10 Myr in galaxies when the outflow H$\alpha$ luminosity contribution is increased, indicating that the feedback somewhat inhibits the star formation within these timescales. The observed emission-line ratios are best reproduced by photoionization models when compared to shock-ionization, indicating that radiation from young stellar population is dominant, and seems to be a consequence of a continuous star-formation activity instead of a bursty event. The outflow properties such as mass outflow rate ($\sim 0.1\,$\myr), outflow kinetic power ($\sim 5.2 \times 10^{-4}\% L_{\mathrm{bol}}$) and mass loading factor ($\sim 0.12$) point towards a scenario where the measured feedback is not strong and has a low impact on the evolution of galaxies in general.
  
  }
   

   \maketitle
%

\section{Introduction}

Galactic-scale outflows are expected to play a major role in regulating star formation \citep{hopkins12,hirschmann13,chisholm17}, mediating the co-evolution of galaxies and their central supermassive black holes \citep{fabian12,king15}, and setting the observed masses of nearby early-type galaxies. There are two main ways of outflows: generated either by the active galactic nuclei (AGN) in galaxy centers, or through powerful winds driven by intense star formation. The galaxy baryonic mass distribution is affected by feedback at low mass (stellar feedback) and at high mass (AGN feedback). In either case, feedback can potentially cause the transfer of large quantities of matter and energy throughout the galaxy interstellar medium (ISM), leading to the enrichment of the ISM \citep{oppenheimer10}, and possibly to the suppression and quenching of star formation \citep{krumholz17,su17}.

Kiloparsec-scale feedback driven by star formation within the ISM is usually described by two physical phenomena: ionization and mechanically-driven outflows. Radiation-pressured stellar winds originate from the intense radiation of young OB stars, which is strong enough to photoionize the ISM gas, mainly in the initial phase of the feedback process, whilst the outflow velocity builds up \citep{chevalier97,veilleux05,girichidis20}. Mechanically-driven outflows then dominate after enough mass is dragged by the feedback, resulting in high-velocity shocks \citep{sharp10}. Supernova events are usually related with the latter type of feedback, displaying characteristic shock-driven ionized gas spectra. Studies of star-formation-driven outflows using integral field spectroscopy (IFS) observations only observe the domination of one of these processes, since they usually happen in smaller scales than the spatial resolution and are probably mixed in the resulting observed spectra \citep[e.g.,][]{ho14,rodriguezdelpino19,dagostino19}.

Expected to dominate the feedback effect in the lower-mass end, star formation regulates itself by outflowing and heating the neighboring gas, which is more abundant in regions with higher star-formation rate (SFR). The higher SFR, however, correlates with the mass outflow rate, creating a regulation cycle that prevents star-forming regions to grow in stellar mass. This relation is given by the equation $\dot{M}_{\mathrm{out}} = \eta\,\mathrm{SFR}$, where $\dot{M}_{\mathrm{out}}$ is the mass outflow rate and $\eta$ is the mass loading factor \citep{schaye10,vogelsberger13,somerville15}.  

From this relation, we expect star-formation-driven outflows to be more powerful at redshifts $z \sim 2$, where the SFR presents its cosmological peak \citep{madau14}.
Indeed, both the SFR and outflow velocities seem to be higher in galaxies at higher redshifts than compared to nearby galaxies \citep{sugahara17,davies19}. Outflows in high-$z$ galaxies, however, are more difficult to measure, usually demanding methodologies that result in loss of the spatial resolution, such as stacking of large sections of galaxies \citep{herrera-camus21} or even summing the integrated spectra of multiple galaxies \citep{ginolfi20}. 

In this sense, IFS studies of low-redshift galaxies present the advantage of a more detailed information in the outflow component for even not so extreme star-formation activity \citep[e.g.,][]{rodriguezdelpino19,zaragozacardiel20}. With the emergence of large IFS surveys such as CALIFA \citep{sanchez12}, SAMI \citep{croom12} and MaNGA \citep{bundy15}, in addition to powerful instruments mounted on large light-collecting telescopes such as VLT/MUSE \citep{bacon10}, GTC/MEGARA \citep{carrasco18} and Gemini/GMOS \citep{allington-smith02}, statistical studies can now be performed in spatial resolutions within sub-kpc scales. This allows us to investigate feedback properties, such as ionization mechanisms, outflow extent, and kinematics, within a large number of galaxies in local scales.

Aside from benefit of higher resolution probing smaller physical scales, low-redshift studies also allow us to trace the bulk of the feedback phenomena in normal star-forming systems instead of those with extreme starburst activity seen at higher redshift. In fact, starburst galaxies represent only 1\% of low-reshift star-forming galaxies, accounting for only $3-6$\% of the stellar production \citep{bergvall16}. Even when we move to higher redshifts such as $\sim 2$, starburst galaxies are only 2\% of the galaxies forming stars, and contribute to 10\% of the SFR density \citep{rodighiero11}. These indicate that star-forming galaxies that experience feedback processes in their cosmic evolution are usually located within (or close to) the star-formation main-sequence \citep{brinchmann04,peng10}.

Characterization of star-formation outflows is a difficult task due to several aspects. Aside from the SFR dependence with redshift, outflow velocities are also dependent on other parameters such as stellar mass. In addition, works in the literature use different methodologies in order to trace the outflow components such as non-parametric \citep[e.g.][]{heckman15}, one-component \citep[e.g.][]{du16} and two-component \citep[e.g.][]{rubin14} measurements, and the use of line ratio maps \citep[e.g.][]{veilleux02} in neutral \citep[e.g.][]{cazzoli14,concas19}, molecular \citep[e.g.][]{leaman19,salak20}, and ionized \citep[e.g.][]{arribas14,robertsborsani20} gas phases. Comparing observational results among each other and with theoretical predictions is not simple, and more information in all these aspects of the outflowing gas is needed to better understand the role of stellar feedback in galaxy evolution.

In this work, we aim to probe the stellar feedback characteristics in a sample of low-redshift star-forming galaxies using spatially-resolved MUSE data. The paper is structured as follows: in Section \ref{sec:data} we present our data, how we selected our sample and detail our methodology. Section \ref{sec:resul} presents the results we obtain by measuring the outflowing gas and the characterization of its parameters. We estimate the outflow properties and energetics in Section \ref{sec:out}. In Section \ref{sec:disc} we discuss our results in comparison with others from the literature, using the estimated outflow energetics. Finally, we conclude in Section \ref{sec:conc}. Throughout this paper we use a comsology with $H_0 = 70.5\,$\kms\,Mpc$^{-1}$, $\Omega_\Lambda = 0.73$ and $\Omega_M = 0.27$ and a \citet{chabrier03} initial mass function (IMF).


\section{Data and Analysis}
\label{sec:data}

In order to probe the properties of star formation and outflows, we have selected a sample of 15 star-forming galaxies at $0.03 < z < 0.2$, which were originally part of the Valparaíso ALMA/APEX Line Emission Survey \citep[VALES;][]{villanueva17,hughes17a} sample of 91 galaxies at $0.02 < z < 0.35$. The final sample was selected following several criteria, described below.

\subsection{The VALES sample}

The VALES sample has been obtained from the {\it Herschel} Astrophysical Terahertz Large Area Survey \citep[{\it H}-ATLAS;][]{eales10}, which provides far-IR observations of extragalactic sky with the PACS and SPIRE cameras in the 100, 160, 250, 350, and 500 $\mathrm{\mu}$m bands \citep{ibar15}. The first 67 galaxies of the VALES sample were selected from the three equatorial fields covered by {\it H}-ATLAS (totaling $\sim 160\,\mathrm{deg}^2$), satisfying the following criteria: (1) flux of $S_{160\mathrm{\mu m}} > 150\,$mJy, close to where typical star-forming galaxies have their flux peaks; (2) sources with no $S_{160\mathrm{\mu m}} > 160\,$mJy ($3\sigma$) detections within 2 arcmin; (3) having an unambiguous identification in the Sloan Digital Sky Survey \citep[SDSS;][]{abazajian09}, with \textsc{reliability}$> 0.8$ \citep{bourne16}; (4) Petrosian SDSS radii smaller than $15''$; (5) high-quality spectra from the Galaxy and Mass Assembly survey \citep[GAMA; \textsc{z\_qual}$\geq 3$;][]{liske15,driver16}. ALMA CO($1-0$) line observations (as part of the VALES collaboration) and GAMA FUV to FIR/submm photometry are also available for these galaxies \citep{villanueva17}. A total of 24 additional starburst galaxies complemented the initial VALES sample observed with the Atacama Pathfinder Experiment APEX/SEPIA Band-5 targeting the CO($J = 2-1$) emission line at $z = 0.1-0.2$ \citep{cheng18}, resulting in the final VALES sample of 91 galaxies. 

\subsection{GAMA data}

GAMA multi-wavelength data are ideal for the analysis of the spectral energy distributions (SEDs) of our sample. In this work we make use of 19 broadband photometric datasets from the ultraviolet to the far-IR included in the GAMA Panchromatic Data Release \citep{driver16} from several facilities: FUV and NUV bands from GALaxy Evolution eXplorer ({\it GALEX}); {\it u}, {\it g}, {\it r}, {\it i} and {\it z} optical bands from SDSS; {\it Y}, {\it J} and {\it H} near-IR bands from the Visible and Infrared Telescope (VISTA); {\it W}1, {\it W}2, {\it W}3 and {\it W}4 mid-IR bands from the {\it Wide-field Infrared Survey Explorer} ({\it WISE}); and 100, 160, 250, 350 and 500 $\mathrm{\mu}$m far-IR bands from H-ATLAS, resulting in SED photometry between 0.15 to 500 $\mathrm{\mu}$m available for our sample.

\subsection{MUSE data}
\label{sec:muse}

Optical IFS data were obtained with the MUSE instrument \citep[Multi Unit Spectroscopic Explorer;][]{bacon10}, mounted on the Very Large Telescope (VLT) for 18 galaxies of the original 91 from the VALES sample as part of the observing program 0100.B-0764(A) (PI Hughes). These galaxies were chosen because of the possibility of tracing the CO(1-0) emission line with ALMA at higher resolution. Observations were performed between November 2017 and February 2018 in dark time. The wide-field mode was used with a field-of-view of $\sim 1^{\prime} \times1^{\prime}$ and a $0.2^{\prime\prime}$ spatial sampling. The wavelength range covered in each datacube is $\sim4750 - 9350$\AA\, with $1.25$\AA\, spectral sampling and a Line Spread Function (LSF) with a full width at half maximum of $\sim 2.5$\AA\, around H$\alpha$ ($\sim 48.5\,$\kms). The spectral resolution ranges from $1750$ in the blue domain to $3750$ in the red \citep{bacon17}.

Each observation was split into five Detector Integration Time (DIT) exposures with offset $90^\circ$ angles. We measured seeing-limited spatial resolutions of $0.58-0.94$ arcsec. These represent physical scales of $0.4-3.3$ kpc at the galaxies, with a median of $1.1$ kpc. The per-pixel signal-to-noise ratio (SNR) at $5500$\AA\, was found to be well above $5$ in the inner disks of the galaxies. Peak SNR values range from $10-373$ with a median of $41$. The quality of the data ensures that we can extract reliable properties for our science goals.

The data reduction pipeline \citep[v2.4.1][]{weilbacher16} was used to process the raw data into science-ready data cubes. This includes bias and dark-frame removal, flat-field corrections, wavelength and flux calibration. For the astrometric calibration and telluric line subtraction we used the default configuration files of the pipeline. Sky exposures near each target were used to correct for the sky continuum. Finally the five DITs were aligned using in-field point sources and combined into a 3D datacube. We assessed the quality of the reduced data by comparing nuclear spectra with those from the GAMA database and found good correspondence in spectral shape and absolute flux.

\subsection{SED modeling with \cigale}
\label{sec:cigale}

\begin{figure*}
\centering
\includegraphics[width=\textwidth]{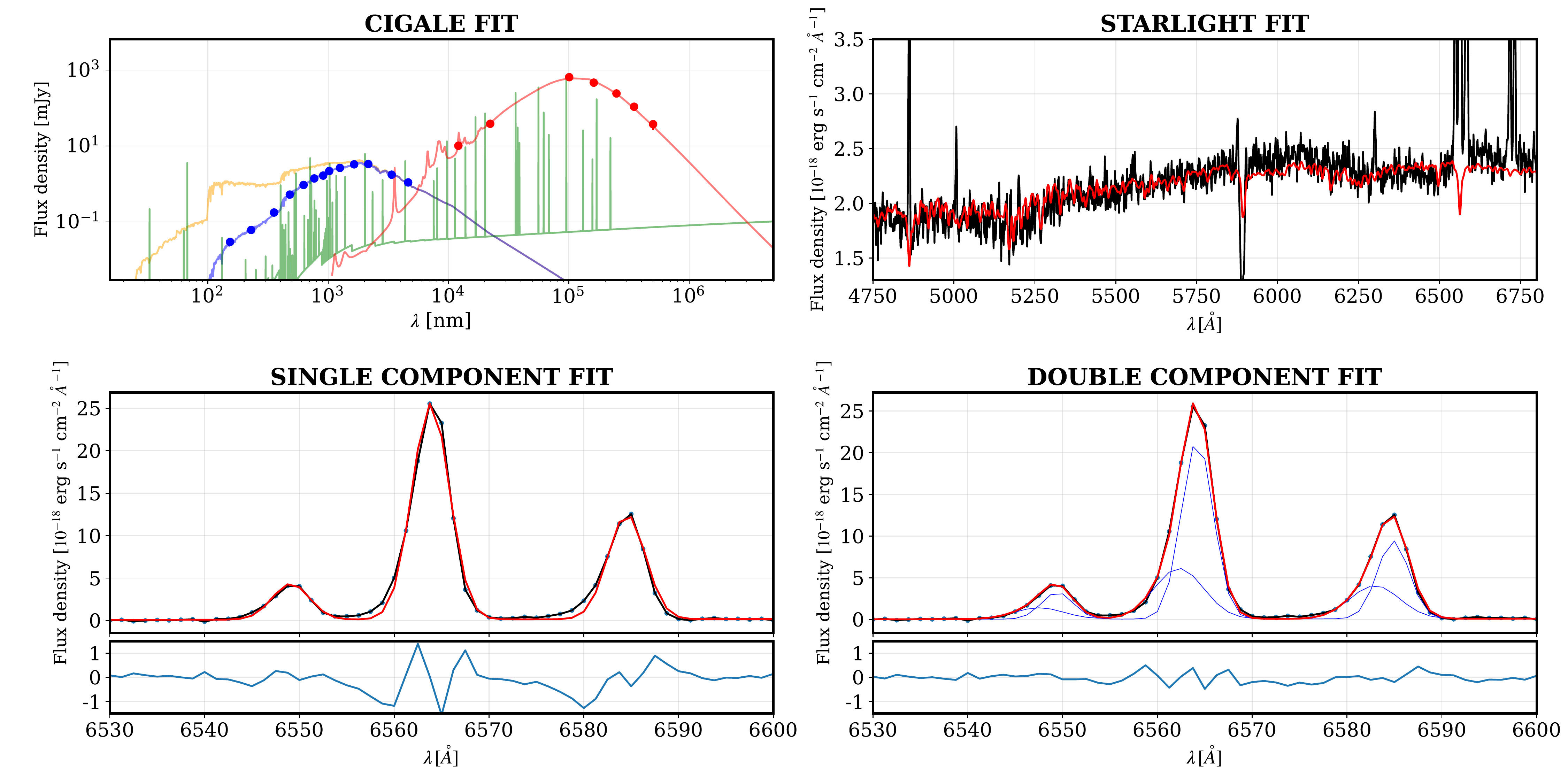}
\caption{Example of spectral fittings for one galaxy (HATLASJ083832). Top left: \cigale\, SED fit, where points represent the broadband magnitudes from the GAMA survey and the red, orange, blue and green lines represent the dust, stellar, stellar atennuated by the dust and emission-line continuum spectra, respectively. Blue points represent the UV and optical bands, and the red points represent the IR bands. Top right: \starlight \, stellar continuum fit, with the black and red lines representing observed and modeled spectra, respectively. Bottom left: single-Gaussian fit of the H$\alpha$+[N\,{\sc ii}] emission lines, where the points and the black line represent the observed spectrum, and the red line represents the modelled lines. A residual spectrum is shown in the bottom part of the panel. Bottom right: same as the previous panel, but showing the double-Gaussian fit, where both components are represented by blue lines. The Gaussian and \starlight\, fits are from the same spaxel, located in the central region of the galaxy.}
\label{fig:fit}
\end{figure*}

\begin{table*}
\caption{Summary of the model parameters used in the SED fitting.}
	\label{tab:sed}
	
\resizebox{\textwidth}{!}{%
    \begin{tabular}{ccccc}  \toprule

Property & Module & Free Parameters & Parameter Description & Values \\ \midrule

              &                  & $q_{\mathrm{PAH}}$  & Mass fraction of polycyclic & 0.47, 1.12, 1.77, 2.50, 3.19, \\ 
              &                  &       & aromatic hydrocarbon (PAH)  & 3.90, 4.58, 5.26, 5.95, 6.63, 7.32 \\ \cmidrule{3-5} 
Dust emission & \citet{draine14} & $U_{\mathrm{min}}$  & Minimum radiation field & [0.1, 50.0] in 36 steps \\ \cmidrule{3-5} 
              &                  & $\alpha$ & Powerlaw slope & [1.0, 3.0] in steps of 0.1 \\ \cmidrule{3-5} 
              &                  & $\gamma$ & Fraction illuminated from & [0.0001,1.0] in 19 steps \\ 
              &                  &          & $U_{\mathrm{min}}$ to $U_{\mathrm{max}}$ & \\ \midrule
              
              &                  & $\tau_{\mathrm{main}}$  & e-folding time of the & 500, 2000, 4000, \\ 
              &                  &       & main stellar population model  & 6000, 8000, 10000 Myr \\ \cmidrule{3-5} 
              &                  & $t_{\mathrm{main}}$  & Age of the main stellar & 11000 Myr \\
Star-formation & SFR delayed     &                      & population in the galaxy &  \\ \cmidrule{3-5} 
history       & + 2 burst/quench & $t_1$ & Age of the first burst/quench & 100 Myr \\ \cmidrule{3-5} 
              & truncations      & $t_2$ & Age of the second burst/quench & 10 Myr \\ \cmidrule{3-5} 
              &                  & $r_1$ & Ratio of the SFR after/before $t_1$ & 0.01, 0.1, 0.5, 1, 2, 5, 10 \\ \cmidrule{3-5} 
              &                  & $r_2$ & Ratio of the SFR after/before $t_2$ & 0.1, 0.5, 1, 2, 5, 10, 20, 50 \\ \midrule

Stellar       & \citet{bruzual03}&   IMF     & Initial mass function & \citet{chabrier03} \\ \cmidrule{3-5} 
populations   &                  & $Z$ & Metallicity & 0.008 \\ \midrule

              &                  & E(B-V) lines  & Nebular emission colour excess & [0.1, 1] in steps of 0.05 \\ \cmidrule{3-5}
              &                  & E(B-V) factor  & Reduction factor to  & 0.25, 0.5, 0.75 \\
Dust          & Modified starburst &              & compute stellar E(B-V)  &                 \\ \cmidrule{3-5} 
attenuation   & attenuation law  & $\delta$ & Attenuation law slope  & [-0.8, 0.0] in steps of 0.1 \\ \cmidrule{3-5} 
              &                  & $R_v$ & Ratio of total to & 3.1 \\ 
              &                  &          & selective extinction& \\ \midrule

    \end{tabular}}
\end{table*}

To reproduce the SED profiles of our sample with GAMA data, we make use of the \cigale\, code \citep[Code Investigating GALaxy Emission\footnote{https://cigale.lam.fr/};][]{burgarella05,noll09,boquien19}. \cigale\, computes the star-formation history (SFH) of a galaxy and combines it to single stellar population models, estimates the emission from ionized gas and applies flux attenuation due to dust. Dust emission based on energy balance in the mid- and far-IR is also taken into account. This is added to possible AGN and radio synchrotron emission to model the multi-wavelength data given as an input to the code. As a result, \cigale\, gives many physical parameters obtained from the fit, such as SFHs, stellar mass, dust luminosities and many others. One can inspect the quality of the fit through evaluation of the best \chisquare\, and reduced \chisquare\, as $\chi^2_r = \chi^2 / (N-1)$, where $N$ is the number of bands fitted. Aside from the best fit obtained through $\chi^2$ minimization, \cigale\, also performs a Bayesian estimate of the physical properties through a likelihood-weighted mean, computed for all grid models, and a 1$\sigma$ uncertainty represented by the peak width of the probability distribution function. The values used from the SED fits in this paper were obtained from the Bayesian approach, unless stated otherwise, as more reliable estimates should be obtained \citep[see][for a more detailed discussion]{noll09}. Mock galaxies catalogues can also be created to estimate the robustness of the fit, where \cigale\, computes the flux densities from the fit added with instrumental noise, and then re-analyzes this mock data to obtain the physical parameters and compares with the results of the original fit. The main goal of the SED fitting in this work is to obtain SFHs and other spatially-integrated parameters for our sample and use them to compare with possible outflowing scenarios in the sample. The top left panel of Fig.~\ref{fig:fit} displays an example of \cigale\, fit for one galaxy of our sample. 

We used the following GAMA broad bands in the SED fit: GALEX FUV and NUV; SDSS u, g, r, i and z; VISTA VIRCAM Y, J, H and Ks; WISE1, 2, 3 and 4; Herschel PACS100, PACS160, SPIRE 250, 250 and 500$\mu$m. We also included the H$\beta$ and H$\alpha$ integrated fluxes to the fit, obtained from the single component fit with MUSE data. Since we are more interested in the SFH parameters of the fit, we decided to first fit the dust contribution of the SED distribution, which is a source of uncertainty in the fit, since the infrared bands present considerably larger uncertainties in comparison to the other bands. We then fitted the infrared bands (from WISE3 to longer wavelengths) isolated just with the \citet{draine14} models to fit the dust contribution with four free parameters. 

\begin{figure}
\centering
\includegraphics[width=\columnwidth]{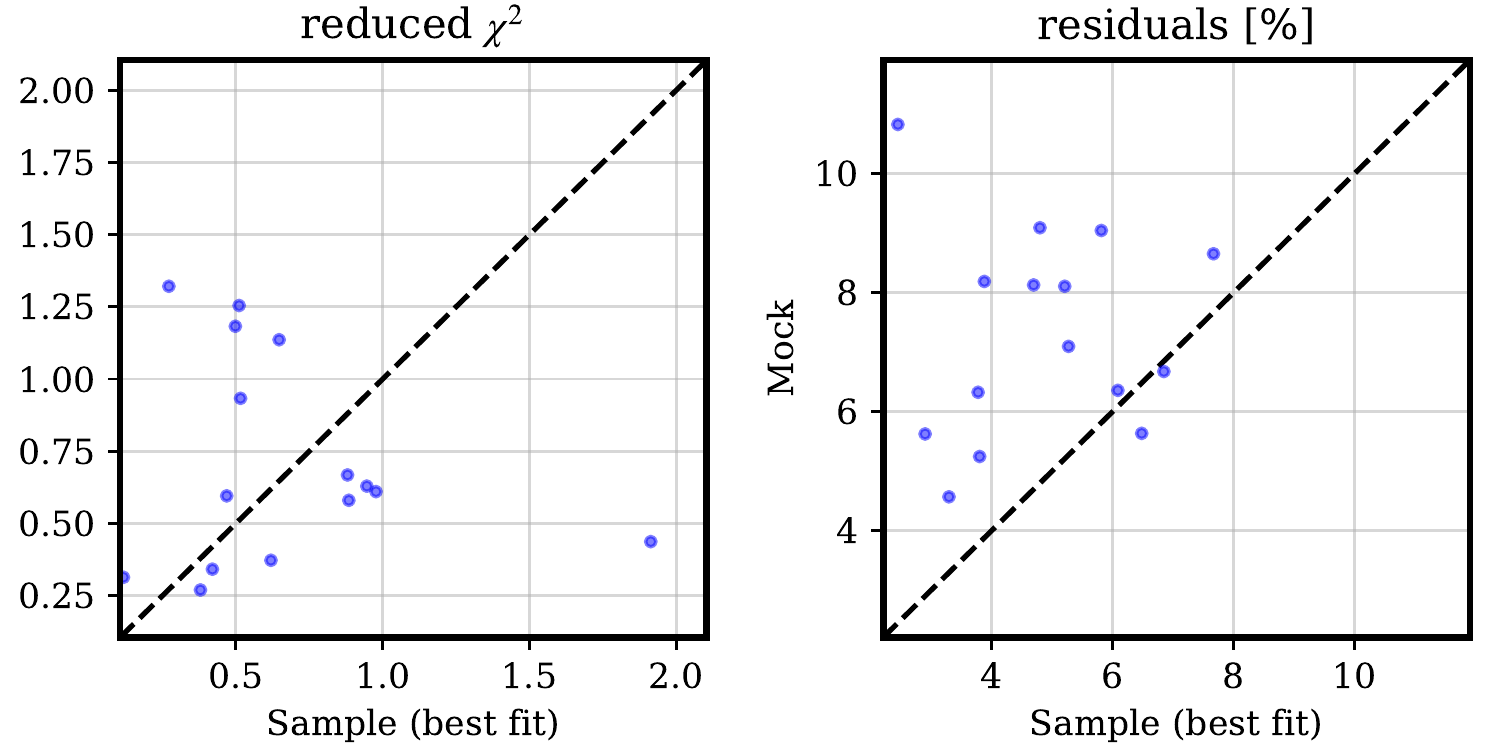}
\caption{Quality of the SED modeling of the lower-wavelength photometric broadbands (from FUV to WISE2), with the addition of the H$\alpha$ and H$\beta$ integrated fluxes from MUSE data. Total reduced chi-square and median residuals (in percentage of the observed magnitudes from GAMA and MUSE), for both the estimated for the VALES sample and the exact for the mock sample, are displayed in the left and right panels, respectively.}
\label{fig:mock}
\end{figure}

We then used the resulting dust luminosities to fit the other bands including the stellar \citep{bruzual03} and nebular contributions, the dust attenuation \citep[modified attenuation law from][with two free parameters]{charlot00} and the SFH function. We assumed a delayed exponential SFH with two constant burst/quenching terms, as described by the equation:

\begin{equation}
\text{SFR}(t) = \left\{
\begin{array}{lr}
t \, \times \, \text{exp}(-t/\tau) & \text{if $t \leq t_1$};\\
r_1 \, \times \, \text{SFR}(t_1) & \text{if $t_1 < t \leq t_2$};\\
r_2 \, \times \, \text{SFR}(t_2) & \text{if $t > t_2$,}
\end{array}
\right .
\end{equation} 

\noindent
where $r_1$ and $r_2$ are the constant burst/quenching terms for two recent time periods $t_1$ and $t_2$. Assuming we can constrain the recent SFR, since we include H$\beta$ and H$\alpha$ emission lines together with the UV bands in the fitting procedure, we fixed $t_1 = 100\,$Myr and $t_2 = 10\,$Myr ago (in lookback time). Aside from $r_1$ and $r_2$, we also left $\tau$ as a free parameter. A summary of the parameters used in the fit is described in Table~\ref{tab:sed} and the resulting SED fit for each galaxy is shown in the Appendix \ref{app:gal_maps}. We also show the reduced chi-square and residuals of the UV and optical part of the SED fit for the VALES sample fit in comparison with the mock models in Fig.~\ref{fig:mock}. As showed, the best-fit results for the VALES sample are comparable with the mock models results, indicating quality sample fits.

\subsection{Stellar population synthesis with \starlight}

In order to obtain ionized emission-line gas properties, we must first remove the stellar continuum and absorption contribution to the IFS spectra. This was performed using the \starlight\, spectral synthesis code \citep{cid05}. \starlight\, fits the observed spectrum with a combination of single stellar populations, with different ages and metallicities, assuming a single dust screen with $R_V = 3.1$, following \citet{cardelli89} dust law. Before we apply the code to the MUSE datacubes, we have corrected the spectra for redshift and Galactic extinction using the \pycasso\, software \citep{cid13,deamorim17}. We used a base of 45 simple stellar-population models from \citet{bruzual03} with a \citet{chabrier03} initial mass function, STELIB stellar library \citep{leborgne03} and `Padova (1994)' evolutionary tracks \citep{alongi93,bressan93,fagotto94a,fagotto94b,fagotto94c,girardi96}. The base of templates present 15 age bins between 1 Myr and 13 Gyr and 3 metallicity bins, $Z = 0.004$, 0.02 and 0.05. The top right panel of Fig.~\ref{fig:fit} shows a \starlight \, fit example for one spaxel of one of the galaxies in the sample.

\subsection{Emission-line Measurements}
\label{sec:el_fit}

Emission-line parameters such as integrated flux, centroid velocity and velocity dispersion were obtained by fitting Gaussian profiles to the stellar-continuum free residual spectra resulted from the \starlight\,fittings to the MUSE datacubes. A fitting code developed by our group\footnote{An early development version of the fitting code can be found at https://github.com/gscouto/demon}, which makes use of the \lmfit\,package \citep{newville14}, a non-linear least-square minimization solver based on the Levenberg-Marquardt algorithm from \scipy\, \citep{virtanen20}, was used to fit the emission lines. The main emission lines (H$\beta$, [O\,{\sc iii}]$\lambda\lambda$4959,5007, [O\,{\sc i}]$\lambda\lambda$6300,6366, H$\alpha$, [N\,{\sc ii}]$\lambda\lambda$6548,6584 and [S\,{\sc ii}]$\lambda\lambda$6717,6731) were fitted. We constrain the line central wavelength and width to be the same for the same ionic species ([N\,{\sc ii}] and H$\alpha$ lines would also share these parameters) and the [O\,{\sc iii}]$\lambda$5007/4959, [O\,{\sc i}]$\lambda$6300/6366 [N\,{\sc ii}]$\lambda$6584/6548 ratios are fixed to the values predicted by atomic physics. Before we applied the emission-line fittings, we employed the Voronoi tesselation technique \citep{cappellari03} with a target S/N ratio of 10 at 6400-6500\AA\, rest-frame spectral region.

We fitted the datacubes with two strategies: (i) one Gaussian profile for each emission line; (ii) initially one Gaussian for each emission line, but then selecting spaxels where a second component would be necessary. In the latter case, a second, broader, component would be added, and the selected spaxels would be refitted. The criterion selected to flag a spaxel for the double component fit is for two consecutive spectral pixels, within the Gaussian profiles (not within the fitted continuum), to be above or below the 5$\sigma$ range, where $\sigma$ is the noise surrounding the fitted emission line. Finally, we created map masks for ``good'' spatial pixels using the following criteria, which were applied to the single component H$\alpha$ fit: (i) flux uncertainties should be below 20$\%$ of the measured integrated flux; (ii) peak velocity and velocity dispersion uncertainties should be below $10\,$\kms. Uncertainties were obtained through \lmfit\, fitting, which explores the parameter space, estimating confidence intervals for variable parameters. Regions that would not fulfill the above criteria were masked out. We also apply the line fitting accounting for the MUSE LSF such as $\sigma = (\sigma_{\mathrm{obs}}^2 - \sigma_{\mathrm{LSF}}^2)^{1/2}$, using the instrumental dispersion values described in Sec. \ref{sec:muse}. No corrections regarding the point-spread function (PSF) was taken into account when fitting the emission lines. Fig.~\ref{fig:fit} shows an example of the two emission-line fitting strategies.

We inspected the possibility of the second component arising due to beam smearing due to the spatial resolution limitations. We follow the methodology applied by \citet{gallagher19}, comparing the second component velocity dispersion with the stellar continuum velocity dispersion we obtain with \starlight. If beam smearing effect would to be considerable, the broadening measured by the second component would also be present in the stellar velocity dispersion. We find that $\sigma_{\mathrm{2nd}}$ is usually considerably higher ($\sim 2\times$) than $\sigma_*$ for the Voronoi regions we trace the second component, with median values of $\sigma_{\mathrm{2nd}} = 110\,$\kms and $\sigma_* = 47\,$\kms, and $\sigma_{\mathrm{2nd}}$ exceeding $\sigma_*$ $\sim 92\%$ the 1$\sigma$ level. This indicate that beam smearing has no effect in the broad results of the paper.

We also note that, for the SED fitting described in Sec.~\ref{sec:cigale}, we used the integrated H$\alpha$ and H$\beta$ fluxes summing the first and second components from the double component emission-line fit.

\subsection{Final sample}

From the 18 galaxies observed with MUSE, we have removed two galaxies (HATLASJ090750 and HATLASJ084630) due to being interacting galaxies and one galaxy (HATLASJ085835) due to being an active galaxy. These galaxies were removed due to our emission-line fitting strategy. The first two clearly present more than one narrow component, probably because of superposition of different galaxies along the line of sight \footnote{This could also be true for several galaxies in our sample, as they show signs of interaction as well. However, as one can observe in the individual maps displayed in Appendix \ref{app:gal_maps}, the regions where we detect a second component are usually located within the galaxies' central regions, thus are not due to galaxy superposition.}. The latter galaxy presents clear AGN type I broad hydrogen profiles, thus affecting the measurement of star-forming emission. We do not remove further galaxies from being AGNs: in fact there are a couple of galaxies suggesting the presence of AGNs through diagnostic diagrams (HATLASJ085828 and HATLASJ090532). Regions related to AGN ionization in the BPT diagram \citep{baldwin81} were removed from our analysis. 

\begin{figure}
\centering
\includegraphics[width=0.5\textwidth]{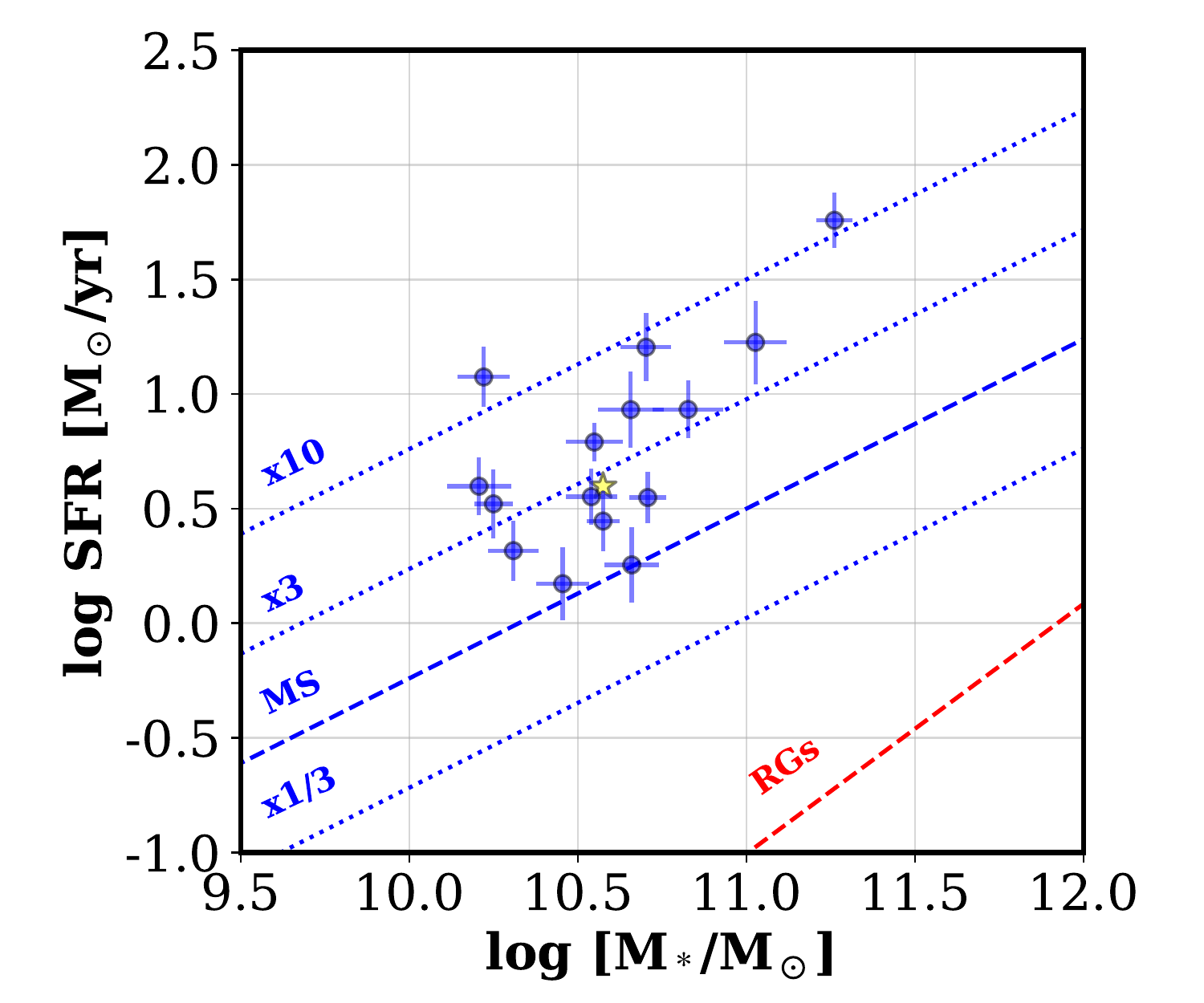}
\caption{Galaxy sample distribution in the M$_*$-SFR diagram (blue dots). The median of the distribution is represented by the yellow star. Values were obtained from the \cigale\,fit of the sample. Blue and red dashed lines represent the star-forming main sequence and retired-galaxy relations obtained by \citet{cano-diaz19}. Dotted lines indicate SFR offsets by the displayed multipliers.}
\label{fig:mass_sfr}
\end{figure}

The final sample consists of 15 galaxies, distributed along the M$_*$-SFR diagram as shown in Fig.~\ref{fig:mass_sfr}, spanning the stellar mass interval of 10.2 < log (M$_*$/M$_\odot$) < 11.3 (median log (M$_*$/M$_\odot$) = 10.6) and 0.2 < log SFR (\myr) < 1.8 \, (median log SFR (\myr) = 0.6). We note that some galaxies in our sample present SFRs of up to 10 times the star-formation main sequence. The sample is, however, generally closer to the main sequence than to starburst values. Further details about each galaxy of the sample are described in the Appendix \ref{app:gal_maps}.

In comparison, the values reported from \citet{villanueva17} for the first 67 VALES galaxies are 0.1 < log SFR (\myr) < 1.9 \, (median log SFR (\myr) = 1.2) and 10.7 < log (M$_*$/M$_\odot$) < 11.3 (median log (M$_*$/M$_\odot$) = 10.7). These parameters were obtained using \magphys\, SED fitting code \citep{dacunha08}. The sample we are probing, therefore, has similar values with respect to the VALES mother sample.

\subsection{Derived spatially-resolved maps}

\begin{table*}
\caption{Summary of the galaxy sample parameters. Stellar masses, SFRs and nebular E(B-V) were obtained through SED fitting, while median gas metallicities were derived using the O3N2 calibrator from the emission-line fit.}
	\label{tab:sample}
	
\resizebox{\textwidth}{!}{%
    \begin{tabular}{ccccccccc}  \toprule

GAMA ID & HATLAS ID & RA (J2000)    & DEC (J2000)   & $z$       & log [M$_*$/M$_\odot$] & SFR [\myr] & $Z$  & E(B-V) \\ \midrule

214184 & HATLASJ083601.5+002617 & 08:36:01.6 & +00:26:18.1 & 0.0332 & $10.46 \pm 0.18$ & $1.44\pm0.55$ & $0.0086\pm0.0020$ & $0.65\pm0.13$ \\ 
208589 & HATLASJ083831.9+000045 & 08:38:31.9 & +00:00:45.0 & 0.0781 & $10.55 \pm 0.19$ & $6.30\pm1.47$ & $0.0086\pm0.0013$ & $0.64\pm0.07$ \\ 
417395 & HATLASJ084217.7+021222 & 08:42:17.9 & +02:12:23.4 & 0.096  & $10.66 \pm 0.22$ & $8.47\pm3.65$ & $0.0081\pm0.0013$ & $0.47\pm0.13$ \\ 
376293 & HATLASJ085111.5+013006 & 08:51:11.4 & +01:30:06.9 & 0.0594 & $10.66 \pm 0.19$ & $1.80\pm0.68$ & $0.0087\pm0.0014$ & $0.70\pm0.12$ \\ 
600024 & HATLASJ085346.4+001252 & 08:53:46.3 & +00:12:52.4 & 0.0504 & $10.31 \pm 0.17$ & $2.07\pm0.62$ & $0.0074\pm0.0019$ & $0.47\pm0.10$ \\ 
600026 & HATLASJ085356.5+001256 & 08:53:56.3 & +00:12:56.3 & 0.0508 & $10.25 \pm 0.13$ & $3.32\pm1.15$ & $0.0087\pm0.0013$ & $0.45\pm0.12$ \\ 
301346 & HATLASJ085406.0+011129 & 08:54:05.9 & +01:11:30.4 & 0.0441 & $10.22 \pm 0.23$ & $4.02\pm1.75$ & $0.0072\pm0.0009$ & $0.32\pm0.15$ \\ 
386720 & HATLASJ085450.2+021207 & 08:54:50.2 & +02:12:08.3 & 0.0583 & $10.71 \pm 0.12$ & $3.54\pm0.91$ & $0.0093\pm0.0016$ & $0.80\pm0.09$ \\ 
622694 & HATLASJ085828.5+003815 & 08:58:28.6 & +00:38:14.8 & 0.0524 & $10.58 \pm 0.11$ & $2.76\pm0.89$ & $0.0093\pm0.0014$ & $0.45\pm0.12$ \\ 
382034 & HATLASJ085957.9+015632 & 08:59:57.9 & +01:56:34.2 & 0.1943 & $10.83 \pm 0.24$ & $8.98\pm2.80$ & $0.0099\pm0.0009$ & $0.76\pm0.11$ \\ 
382362 & HATLASJ090532.6+020220 & 09:05:32.6 & +02:02:21.9 & 0.0519 & $10.54 \pm 0.17$ & $3.57\pm1.00$ & $0.0100\pm0.0016$ & $0.48\pm0.09$ \\ 
324842 & HATLASJ090949.6+014847 & 09:09:49.6 & +01:48:46.0 & 0.1819 & $11.25 \pm 0.13$ & $61.44\pm16.29$ & $0.0084\pm0.0025$ & $0.67\pm0.09$ \\ 
324931 & HATLASJ091157.2+014454 & 09:11:57.2 & +01:44:53.9 & 0.1694 & $11.03 \pm 0.21$ & $16.80\pm7.03$ & $0.0095\pm0.0007$ & $0.52\pm0.13$ \\ 
534896 & HATLASJ114244.3-005450 & 11:42:44.3 & -00:54:48.8 & 0.1076 & $10.22 \pm 0.18$ & $11.99\pm3.89$ & $0.0062\pm0.0011$ & $0.39\pm0.11$ \\ 
319694 & HATLASJ142128.2+014845 & 14:21:28.1 & +01:48:44.3 & 0.1604 & $10.70 \pm 0.17$ & $15.73\pm6.11$ & $0.0073\pm0.0012$ & $0.35\pm0.12$ \\ \midrule

    \end{tabular}}
\end{table*}

Several distribution maps are derived from the methodology described above to aid our analysis. We estimate the SFR surface density ($\Sigma_{\mathrm{SFR}}$) using the dust-corrected H$\alpha$ flux \citep[extinction law from][]{cardelli89} measured summing the first and the second component from the double component emission-line fit. We used the relation log SFR = log $L_{\mathrm{H}\alpha} + k$ \citep{kennicutt12} with the $k$ coefficients from \cite{zezas21}. This coefficient, which we also mapped its spatial distribution, depends on the assumed IMF (Chabrier) and the gas metallicity. Metallicity maps were constructed using the \citet{marino13} relation using the O3N2 calibrator. Specific SFRs, sSFR = $\Sigma_{\mathrm{SFR}}$/$\Sigma_{\mathrm{M*}}$, were also obtained using the stellar-mass surface density resulted from the \starlight\,fit. We constructed $\Sigma_{\mathrm{SFR}}$, $\Sigma_{\mathrm{M*}}$ and sSFR maps for the galaxy sample, which are shown in the individual figures in the Appendix \ref{app:gal_maps}. A summary of the galaxy sample physical properties is described in Table~\ref{tab:sample}.


\section{Results}
\label{sec:resul}

\subsection{Single-component fit}
\label{sec:1compfit}

\begin{figure*}
\centering
\includegraphics[width=\textwidth]{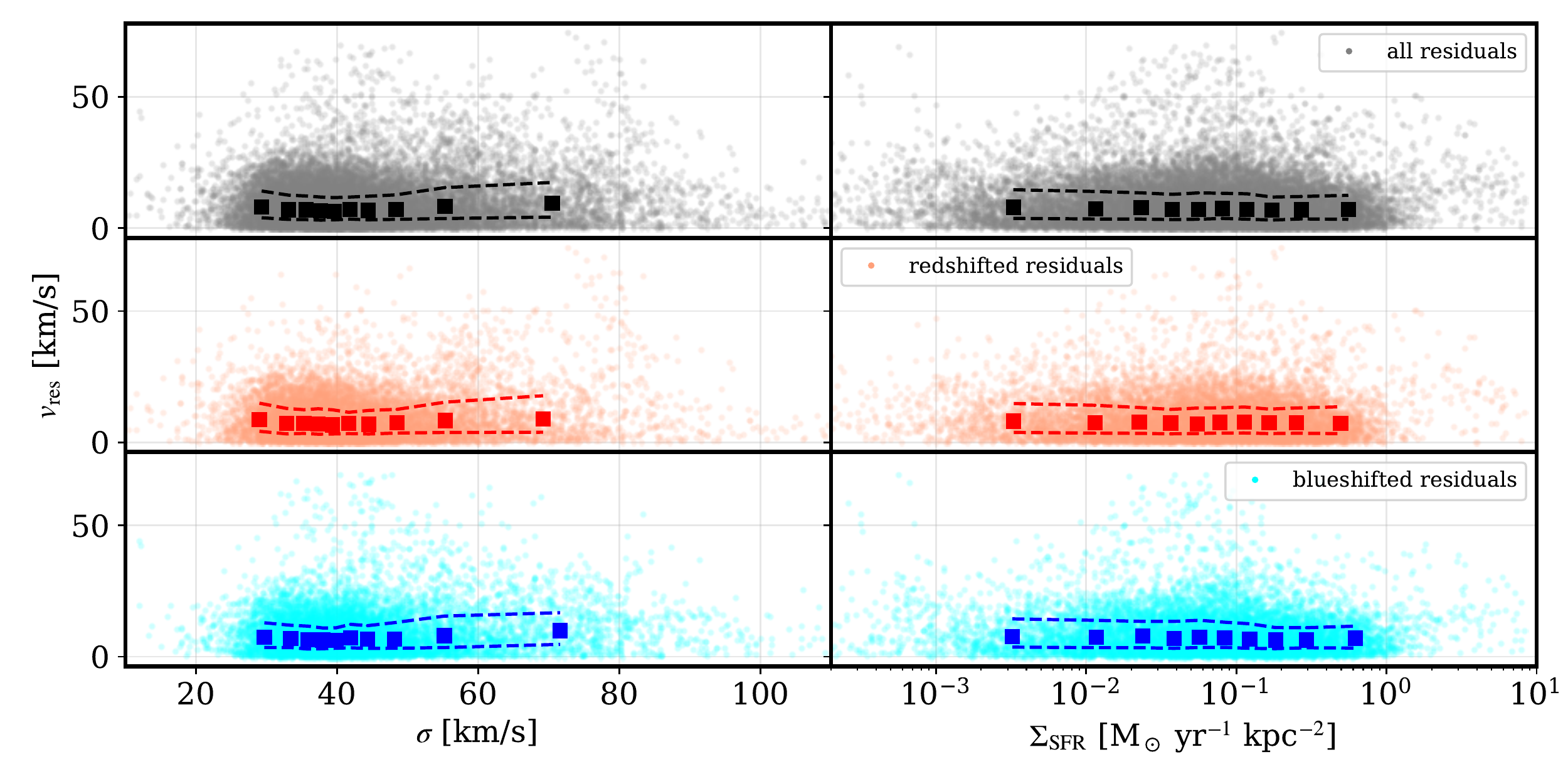}
\caption{Relation between the velocity disperion $\sigma$ (left) and SFR density $\Sigma_{\mathrm{SFR}}$ (right) with the residual velocities obtained from the rotation model fitted to the H$\alpha$ velocity fields. Each point represents a Voronoi region. The top panels show the distribution for the whole sample, while the middle and bottom panels are separated into redshifted and blueshifted residuals, respectively. Square symbols represent median values, with the dashed lines representing 25 and 75 percentiles of the distribution. No clear correlation between the two parameters is observed with $v_{\mathrm res}$.}
\label{fig:sig_sfr_vs_res}
\end{figure*}

We begin investigating signatures of outflowing gas by looking into the single-component emission-line fit of the spatially-resolved MUSE data. Assuming that the presence of a secondary component would drive the single-component fit to present a somewhat higher velocity dispersion and a displaced peak velocity in relation to what should be expected if this second component were not present (as we can observe in the example shown in the bottom panels of Fig.~\ref{fig:fit}), we can expect the kinematic properties of the single-component fit to indicate the presence of the secondary component by somehow measuring this velocity shift and the increase in velocity dispersion. 

To probe this scenario we have fitted a disk model to the gas velocity field. Following \citet{bertola91}, we assume a spherical potential, with the gas following only circular orbits. The observed radial velocity at the position ($R$,$\Psi$) is given by the equation:

\begin{multline}
    v_{\mathrm{model}}(R, \Psi) =  v_{\rm sys} +\\ 
    \frac{A\,R \cos(\Psi - \Psi_0) \sin \theta \cos^p \theta}{
        \left\{  R^2 \left[ \sin^2 (\Psi - \Psi_0) + \cos^2 (\Psi - \Psi_0) \right] 
        + c_0^2\, \cos^2 \theta \right\}^{p/2}
    } ,
\end{multline}

\noindent
where $v_{\rm sys}$ is the systemic velocity (which is close to zero, since we have redshift-corrected the spectra), $A$ is the centroid velocity amplitude, $R$ and $\Psi$ are the radial and angular coordinates of a given pixel in the plane of the sky, $\Psi_0$ is the position angle of the line of nodes, $c_0$ is a concentration parameter (constraining the radius at which the centroid velocity reaches 70 per cent of the amplitude $A$) and $\theta$ is the disc inclination ($\theta = 0^\circ$ for a face-on disc). Finally, the parameter $p$ measures the slope of the rotation curve after reaching the maximum amplitude. This parameter was fixed to $p = 1$, which corresponds to an asymptotically flat rotation curve at large radii. The best fit was obtained also using the \lmfit\, package with a Levenberg-Marquardt least-squares minimization. We used the optical continuum image of each galaxy as weight in order to better reproduce the observed velocity maps. The resulting velocity model maps for each galaxy of the sample are shown in the Appendix \ref{app:gal_maps}, as well as the residual velocities ($v_{\mathrm{res}} = v_{\mathrm{obs}} - v_{\mathrm{model}}$).

Our main interest in this model is to retrieve the residual velocities. These are representative of the centroid velocity deviation due to the presence of a second component in the line profile. Fig.~\ref{fig:sig_sfr_vs_res} shows the relation between the velocity dispersion and the $\Sigma_{\mathrm{SFR}}$ with the residual velocities for the entire sample. Although some scatter is observed, no correlation between the two parameters is observed with the residual velocities, indicating that the hypothesis raised in the beginning of this section does not hold. Figures of these relations for individual galaxies, shown in the Appendix \ref{app:gal_maps}, show that this scenario could be true for some galaxies of our sample. This could be the case of HATLASJ083832, HATLASJ085356 and HATLASJ090949, for example, where higher residual velocities are observed for higher $\Sigma_{\mathrm{SFR}}$ and velocity dispersion. The $\Sigma_{\mathrm{SFR}}$ and $\sigma$ values where $v_{\mathrm{res}}$ increases, however, do not seem to be the same between these galaxies. This, together with the fact that we may not have the presence of outflows in a good part of the sample, may ``wash out'' the outflow signature in Fig.~\ref{fig:sig_sfr_vs_res}. We will explore this scenario further in Sec. \ref{sec:sed}, where we compare these slopes with the SFR obtained through the SED fitting.

\subsection{Double-component fit}

\begin{figure*}
\centering
\includegraphics[width=\textwidth]{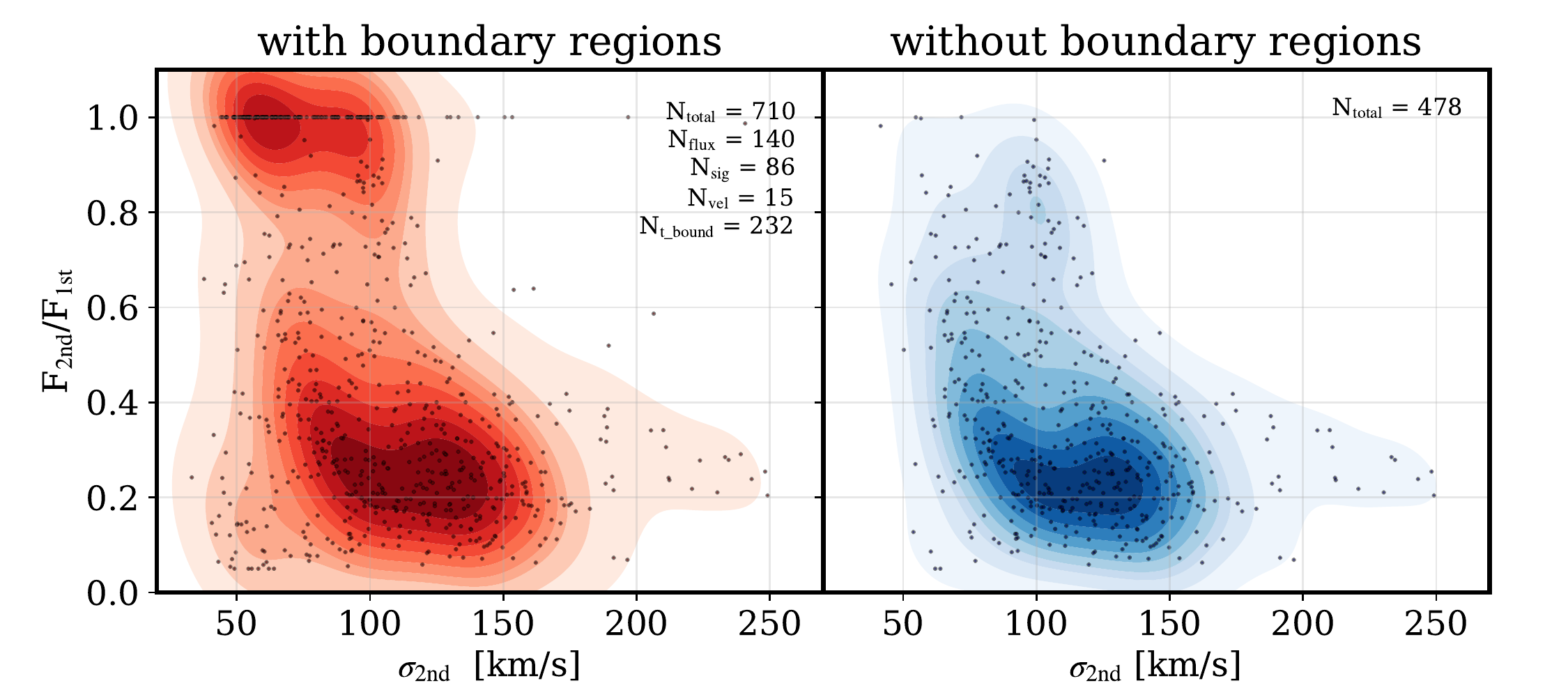}
\caption{Relation between the velocity dispersion of the second component and the flux ratio between the second and the first components, from the double-component fit. The left panel shows all the Voronoi regions where a double-component fit was performed. These include what we call the ``boundary regions'' (explained in the text), and the number of such regions, discriminated by criteria, is shown in the top right. The right panel shows the distribution without boundary regions.}
\label{fig:sig_b_vs_flux_ratio}
\end{figure*}

For regions where a single component did not result in a satisfactory fit, we have added a second component, following the guidelines mentioned in Sec. \ref{sec:el_fit}, considering it to be related with outflowing gas due to star-forming activity. Because it is difficult to obtain strong constraints on its parameters, we have decided to constrain the second component to (i) have lower integrated flux than the first component, so $F_{\mathrm{2nd}} / F_{\mathrm{1st}} < 1$; (ii) its velocity dispersion needs to be higher than $1.2$ and lower than $3.0$ times the first-component velocity dispersion. These values are empirical, and were chosen after testing several velocity dispersion conditions; (iii) its centroid velocity should not differ by more than $200\,$\kms\, (in absolute values) than the one fitted for the first component. We also fixed the H$\alpha$/H$\beta$ and \nii/H$\alpha$ ratios between the first and second components, so both ratios are the same for both components. The Voronoi regions where we fitted a second component represent a small portion of some galaxies in the sample, as can be seen in the flux maps of the second component in the figures shown in the Appendix \ref{app:gal_maps}, for each galaxy in our sample. In fact, we detect the second component in 13 of the 15 galaxies.

\begin{figure*}
\centering
\includegraphics[width=\textwidth]{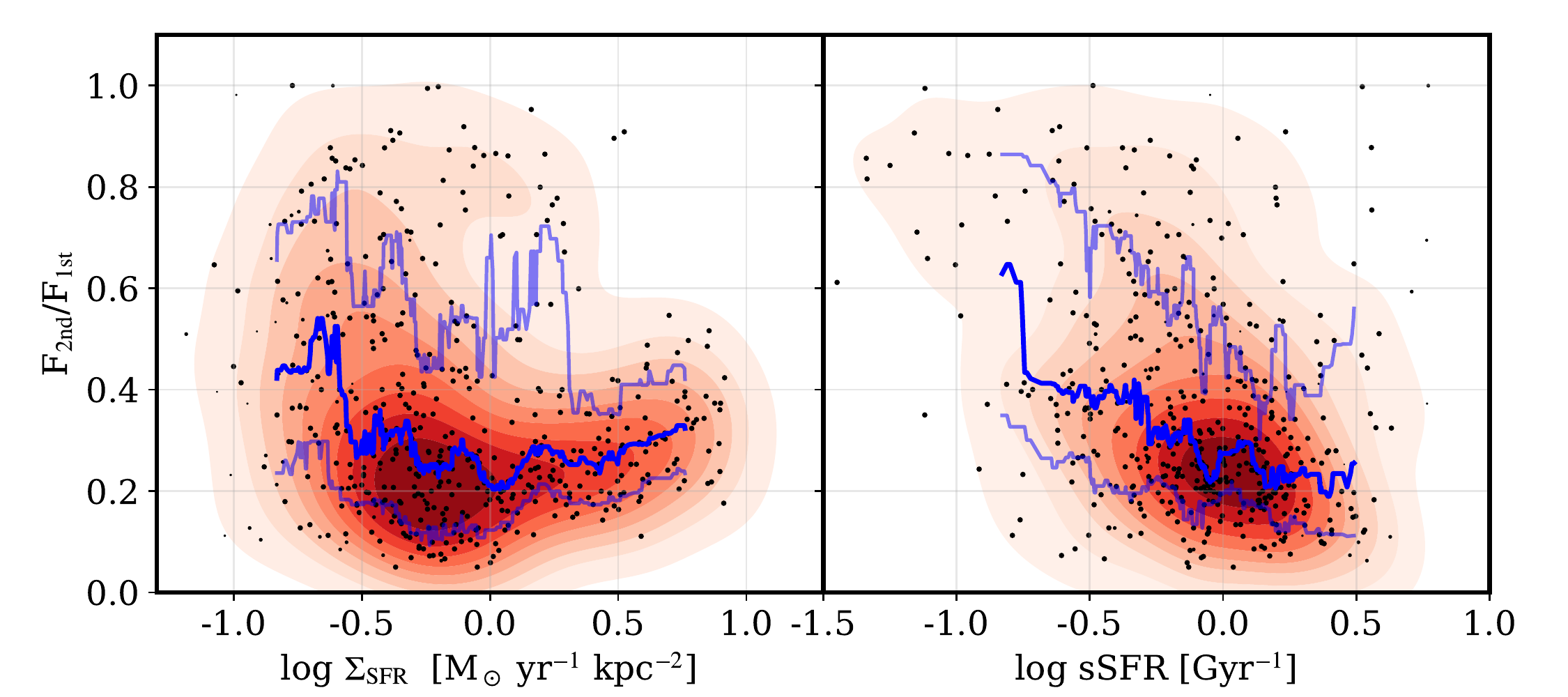}
\caption{Relations between SFR surface density and specific SFRs and second over first components flux ratio. Black points represent Voronoi regions, where the point size is proportional to the inverse of the Voronoi covered area, so small points represent regions where more spaxels had to be binned to reach a satisfactory S/N ratio. Strong blue lines represent a running median with a 20-point window, with weak blue lines representing the 1-$\sigma$ of the distribution.}
\label{fig:sfr_vs_flux_ratio}
\end{figure*}

Fig.~\ref{fig:sig_b_vs_flux_ratio} shows the relation between the velocity dispersion of the second component $\sigma_{\mathrm{2nd}}$ and the integrated H$\alpha$ flux ratio $F_{\mathrm{2nd}}/F_{\mathrm{1st}}$. The left panel includes regions where the emission-line fit results in some parameter boundary (e.g., $F_{\mathrm{2nd}}/F_{\mathrm{1st}} = 1$), which we will call ``boundary regions''. The right panel is the same, removing such regions. From now on, we will only consider none-boundary regions, as we consider these are satisfactorily constrained in our measurement methodology. We can observe in Fig.~\ref{fig:sig_b_vs_flux_ratio} that the bulk of the regions present a $80\,$ \kms$\,< \sigma_{\mathrm{2nd}} < 150\,$\kms\, and a $F_{\mathrm{2nd}}/F_{\mathrm{1st}} \sim 0.2$. The distribution is scattered for lower velocity dispersion, but narrows for values of $\sigma_{\mathrm{2nd}} > 130\,$\kms, where the $F_{\mathrm{2nd}}$ contribution is lower than half of the $F_{\mathrm{1st}}$. Few points reach values $\sigma_{\mathrm{2nd}} > 200\,$\kms, with $0.2 < F_{\mathrm{2nd}}/F_{\mathrm{1st}} < 0.4$. Although there is a suggestion that the $F_{\mathrm{2nd}}/F_{\mathrm{1st}}$ decreases with $\sigma_{\mathrm{2nd}}$ in the $\sigma_{\mathrm{2nd}} > 200\,$\kms\, region, the region is barely populated, and we would need more points to make conclusions for this region.

The left panel of Fig.~\ref{fig:sfr_vs_flux_ratio} shows the relation between $\Sigma_{\mathrm{SFR}}$ and $F_{\mathrm{2nd}}/F_{\mathrm{1st}}$ flux ratio. The $F_{\mathrm{2nd}}/F_{\mathrm{1st}}$ ratio displays scattered distribution up to log $\Sigma_{\mathrm{SFR}} \sim 2.0\,$\myrkpc, but concentrated in values $F_{\mathrm{2nd}}/F_{\mathrm{1st}} < 0.5$. $F_{\mathrm{2nd}}/F_{\mathrm{1st}}$ generally decreases with increasing $\Sigma_{\mathrm{SFR}}$ in this interval. This could be due to that, although the S/N ratio is high overall for all the points in this plot\footnote{We remind that we just apply the second-component fit for regions where the fit residuals from the single component is above the 5$\sigma$ noise threshold. So low S/N regions usually do not qualify for a double-component fit.}, the S/N ratio is probably lower for low $\Sigma_{\mathrm{SFR}}$ values, resulting not only in the scatter observed in the $F_{\mathrm{2nd}}/F_{\mathrm{1st}}$ ratio but also restricting us to trace small $F_{\mathrm{2nd}}$ contributions, and thus not being able to populate low $F_{\mathrm{2nd}}/F_{\mathrm{1st}}$ ratio for low $\Sigma_{\mathrm{SFR}}$ values. This is indicated by the apparent concentration of larger Voronoi regions in the left lower corner of the plot, coming from regions where more binning was needed to reach a satisfactory S/N ratio. The trend shifts with increasing $F_{\mathrm{2nd}}/F_{\mathrm{1st}}$ ratio for log $\Sigma_{\mathrm{SFR}} > 2.0\,$\myrkpc, where the flux ratio is much less scattered, as shown by the 1-$\sigma$ interval lines, indicating a higher flux contribution from the second component in high-SFR regions.

\begin{figure}
\centering
\includegraphics[width=\columnwidth]{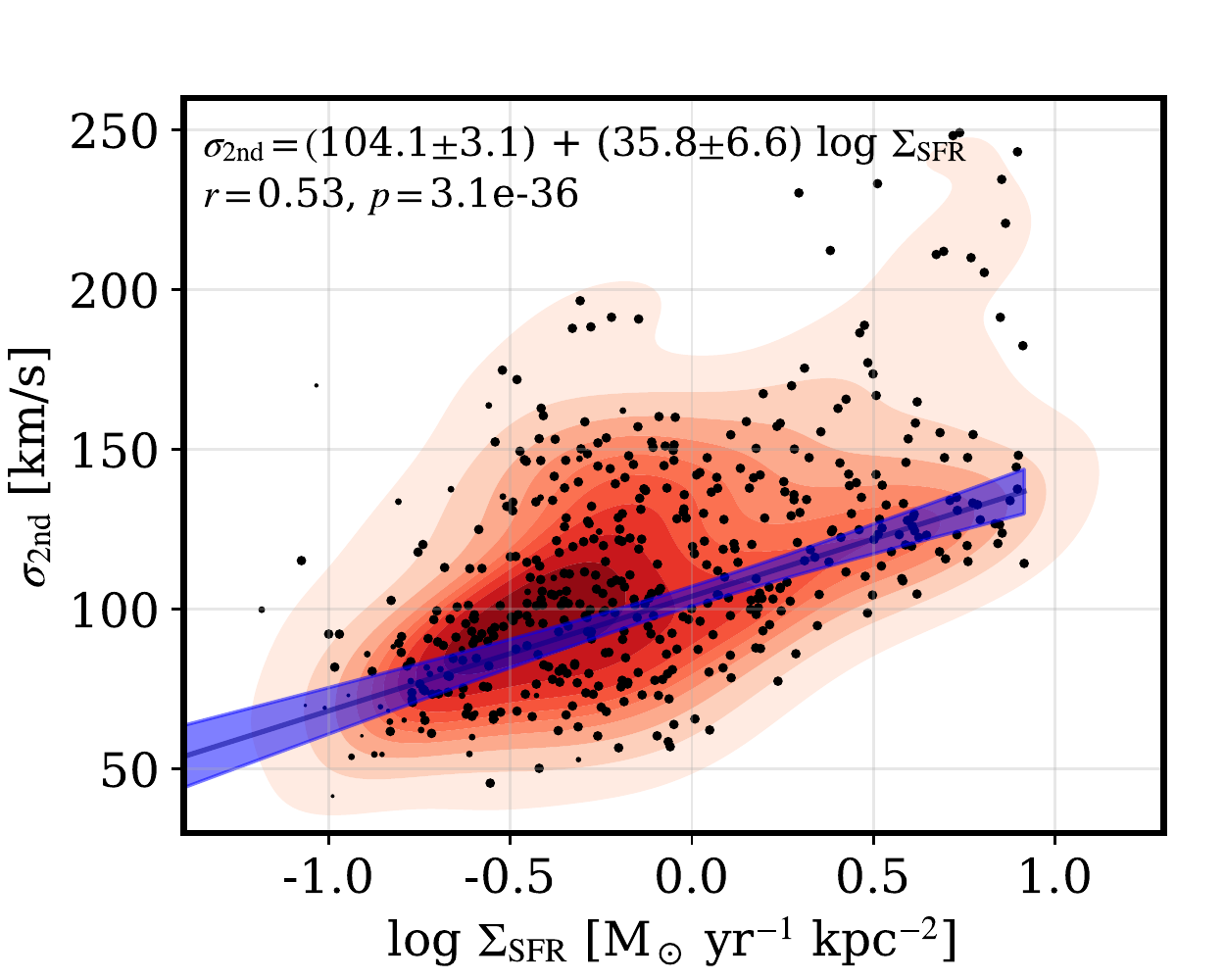}
\caption{Second-component velocity dispersion as a function of the SFR surface density. Voronoi regions are represented by black points, with point size being anticorrelated with the Voronoi region area. The black line and blue shaded region show the best-fit relation with a 3$\sigma$ uncertainty, and the relation is displayed in the top-left corner. Pearson coefficients are also displayed, indicating a significant positive correlation.}
\label{fig:sfr_vs_sigma2}
\end{figure}

\begin{figure*}
\centering
\includegraphics[width=\textwidth]{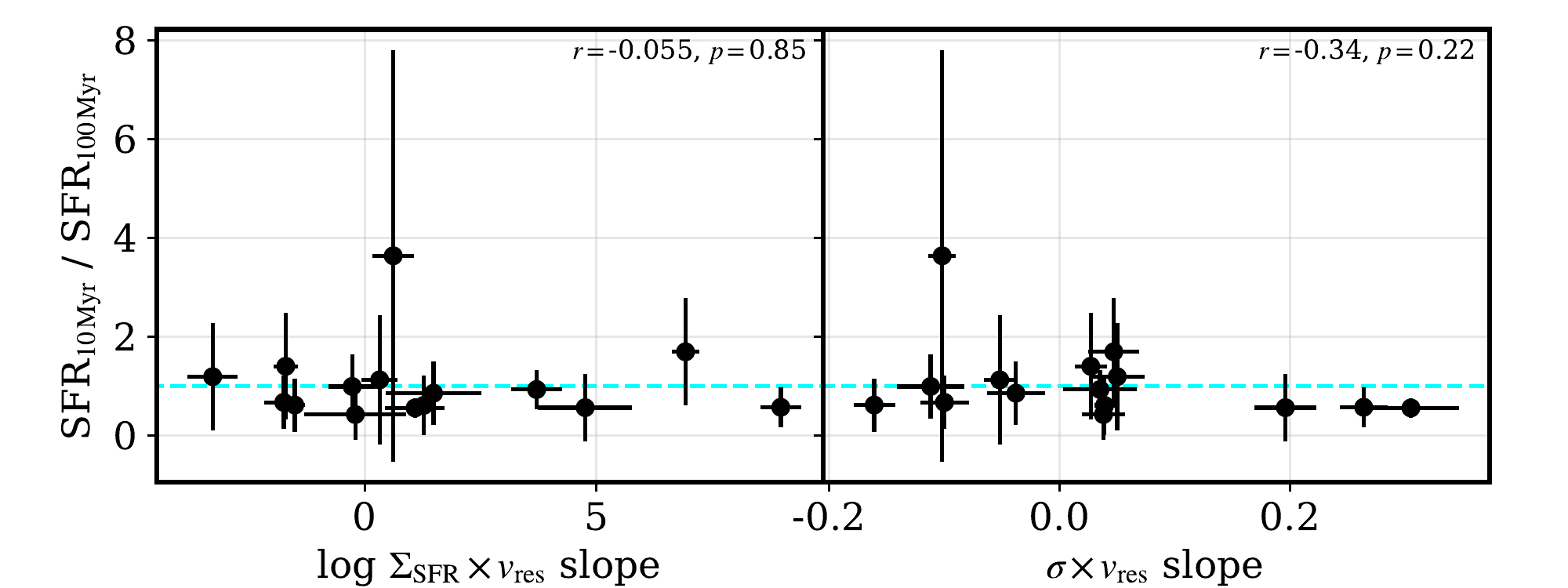}
\caption{Relation between the slopes of log $\Sigma_{\text{SFR}}$ (left panel) and one component-fit $\sigma$ (right) versus residual velocities and the SFR$_{10\,\text{Myr}}$/SFR$_{100\,\text{Myr}}$ ratio obtained with the SED fit. The slopes are obtained through the fits showned for each individual galaxy in the Appendix \ref{app:gal_maps}. Pearson coefficients are shown in the top-right corners, showing no correlation in the first relation, and weak anticorrelation with little significance in the second. The cyan line shows the SFR$_{10\,\text{Myr}}$/SFR$_{100\,\text{Myr}} = 1$ ratio.}
\label{fig:sfr_ratio_slope}
\end{figure*}

The right panel of Fig.~\ref{fig:sfr_vs_flux_ratio} shows the relation between sSFR and $F_{\mathrm{2nd}}/F_{\mathrm{1st}}$ flux ratio, where we observe again a decrease of sSFR and $F_{\mathrm{2nd}}/F_{\mathrm{1st}}$ with a shift in very high values ($\mathrm{log\, sSFR} > 0.3\,\mathrm{Gyr^{-1}}$), although in this case there are fewer points in this trend shift. Again, this could be just dominated by the lower S/N ratio. In this case, there is no clear shift as shown in the left panel (or at least not as clear), where the flux ratio seems to increase after a certain SFR value. Visually inspecting the regions where we trace the second component, we notice these are usually located in galaxy centers, where stellar mass is increased, which could explain the higher $F_{\mathrm{2nd}}$ contribution in lower sSFR values. However, since it is hard to disentangle the dependence factor in this relation, further analysis would be needed to address this relation.

In Fig.~\ref{fig:sfr_vs_sigma2} we display the relation between the $\Sigma_{\mathrm{SFR}}$ and the second-component velocity dispersion, $\sigma_{\mathrm{2nd}}$. Although the $\sigma_{\mathrm{2nd}}$ are not high, concentrating between $80 < \sigma_{\mathrm{2nd}} < 120\,$\kms, we can observe a correlation between the two. This is expected if we consider that the second component is indeed tracing an outflowing gas due to increased star-formation activity. The low $\sigma$ values, however, seem to indicate that these outflows are not strong and of low velocities.

We note that, due to our measurement strategy of fitting a second component in regions where the residuals above the 5$\sigma$ threshold, we expect these regions to present higher signal-to-noise ratios. Thus the fact that we are probing low velocity (and luminosity, as we well see in the following sections) outflows are a result of the deeper observations in this work, when compared to other studies using galaxy surveys (e.g., MaNGA). 

\subsection{SED fit}
\label{sec:sed}

After inspecting the relations in the spatially-resolved MUSE data, now we take a look at the results obtained with the integrated broadband SED photometry fitted with \cigale. In order to probe whether the outflow activity affects the recent bursting or quenching of star formation, we compare the ratio of the SFRs at 10 and 100 Myr obtained from the SED fitting to the physical properties associated with outflowing ionized gas. In Fig.~\ref{fig:sfr_ratio_slope} we compare this ratio with the slopes we observe for each galaxy between $\Sigma_{\text{SFR}}$ and the single-component $\sigma$ with the residual velocities (Appendix \ref{app:gal_maps}). As indicated in Sec. \ref{sec:1compfit}, these properties do not seem to show outflow signature for the entire sample, but could for individual galaxies. However, this does not seem to be the case. We do not observe any correlation between the SFR$_{10\,\text{Myr}}$/SFR$_{100\,\text{Myr}}$ ratio and the $\Sigma_{\text{SFR}}$ and $v_{\text{res}}$ slope, even though some of the galaxies present steep slopes between $v_{\text{res}}$ and $\Sigma_{\text{SFR}}$ (e.g., HATLASJ083832). Although a weak anticorrelation is observed between the $\sigma \times v_{\text{res}}$ slope and the SFR ratio, the significance is too small to take this result into consideration.

This result, in addition to the lack of evidence of outflowing gas in Fig.~\ref{fig:sig_sfr_vs_res} as discussed in Sec.~\ref{sec:1compfit}, indicate that, if gas outflows are present, the simple disk kinematics model we are using here is not detailed enough to account for the non-circular motions these outflows could generate. This is also a clear indication that these outflows would not be strong, since no major kinematics deviation from the galaxies rotating disk is observed. 

\begin{figure}
\centering
\includegraphics[width=\columnwidth]{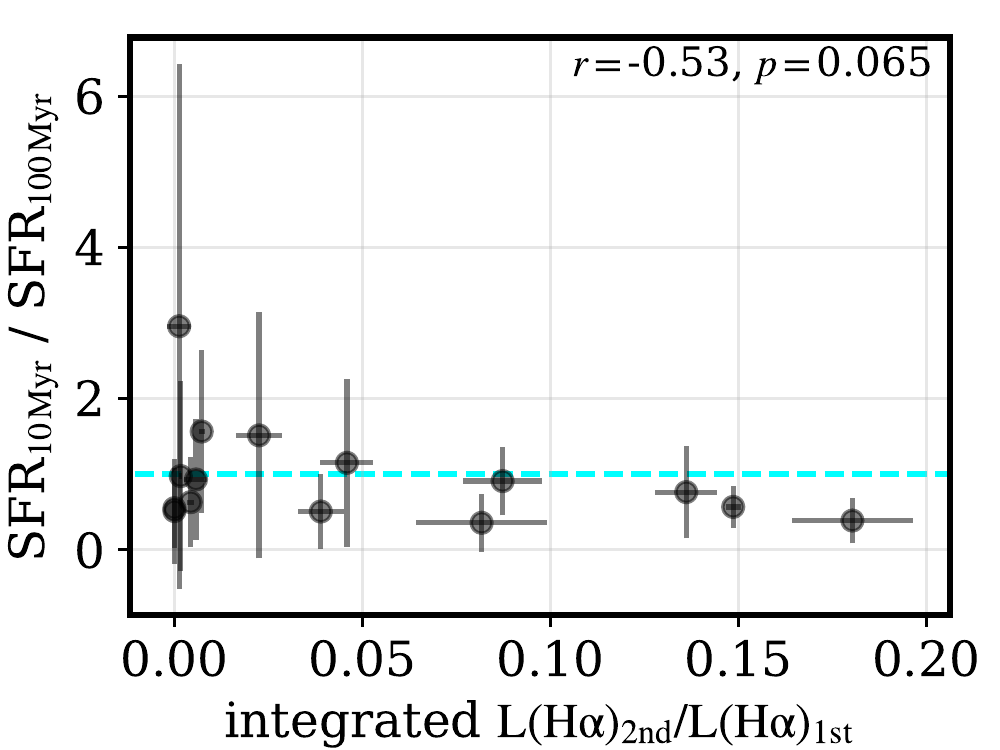}
\caption{Dependence of the SFR$_{10\,\text{Myr}}$/SFR$_{100\,\text{Myr}}$ ratio with the spatially-integrated H$\alpha$ luminosity ratio between the second and the first components. Pearson coefficients are shown in the top-right corner, showing a moderate anticorrelation with a small significance.}
\label{fig:lum_ratio_vs_sfr_ratio}
\end{figure}

Fig.~\ref{fig:lum_ratio_vs_sfr_ratio} displays the dependence of the SFR$_{10\,\text{Myr}}$/SFR$_{100\,\text{Myr}}$ ratio on the integrated H$\alpha$ luminosity ratio between the second and the first components, for the case of the two-component fit. We observe a moderate anticorrelation with a small significance, but higher than the relation observed in the right panel of Fig.~\ref{fig:sfr_ratio_slope}. The significance is still not good enough to consider this anticorrelation to be true, and a larger sample would be needed to confirm this trend. If this is indeed a true anticorrelation, we could be probing the outflow impact timescales in SFR, with the feedback being effective in inhibiting star formation between 10 and 100 Myr. However, even though we could be tracing the outflow signature in the single- and double-component emission-line measurements for some galaxies, the results do not show a clear indication of impact in the recent SFH between 10 and 100 Myr at global scales.

\begin{figure*}
\centering
\includegraphics[width=\textwidth]{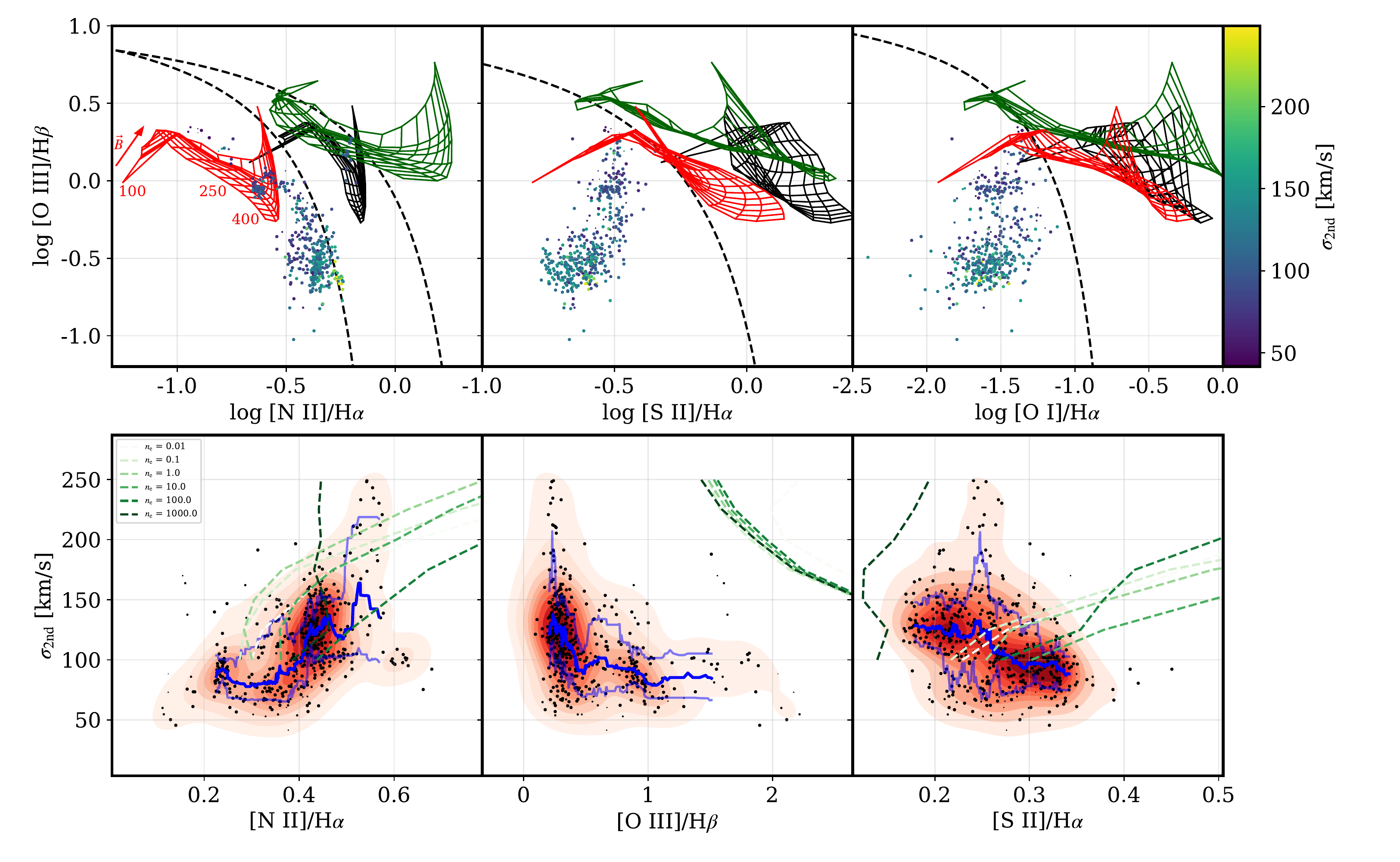}
\caption{BPT diagrams (top panels) and the dependence of the second-component velocity dispersion $\sigma_{\mathrm{2nd}}$ on \nii/H$\alpha$, \oiii/H$\beta$ and \sii/H$\alpha$ line ratios of the Voronoi regions where we do measure a second component related to outflowing gas (bottom panels). The BPT diagrams are colored with its $\sigma_{\mathrm{2nd}}$. Dashed lines display the theoretical \citep{kewley01} and empirical \citep{kauffmann03} regions separating pure star formation and star formation combined with AGN ionization, respectively. Red, black and green grids display \mappings\, shock models for LMC ($Z = 0.0071$), Dopita\_2005 ($Z = 0.0065$) and Solar metallicities ($Z = 0.0183$), respectively. Grids are for electron density $n_{\mathrm{e}} = 1\,$\cmt, with magnetic fields $0.0001 < \vec{B} < 10\,\mu$G and shock velocities $ 100 < v_{\mathrm{shock}} < 400\,$\kms (shock velocities usually vary along the x-axis of the BPT diagram). Some shock velocity values are indicated, as well as the general direction of increasing magnetic field intensity. Blue lines in the bottom panels represent the running median with 1$\sigma$ distribution of the Voronoi regions. Green dashed lines represent Solar metallicity shock models with $\vec{B} = 1\,\mu$G and $0.01 < n_{\mathrm{e}} < 1000\,$\cmt, identified by the labels. Y-axis values for the green dashed lines are $v_{\mathrm{shock}}$.}
\label{fig:bpt_diagram}
\end{figure*}

We note that for five of the galaxies in our sample, companions can be observed within the MUSE FoV, and these have different GAMA IDs. This means that the GAMA data used to do the SED modelling do not integrate the entire MUSE FoV emission, which could lead to discrepancies when comparing the emission-line and SED results, the main topic of this section. Thus, for Figs. \ref{fig:sfr_ratio_slope} and \ref{fig:lum_ratio_vs_sfr_ratio} we have estimated the slopes and the luminosity ratios masking out the companion galaxies. These companion galaxies, however, have negligible contribution to the $v_{\text{res}}$ slopes and L(H$\alpha$)$_{\mathrm{2nd/1st}}$ ratio, and the masking does not change the results obtained here. These galaxies are HATLASJ084217, HATLASJ085406, HATLASJ090949, HATLASJ114244 and HATLASJ142128.


\subsection{Gas excitation}
\label{sec:gas_exc}

\begin{figure*}
\centering
\includegraphics[width=\textwidth]{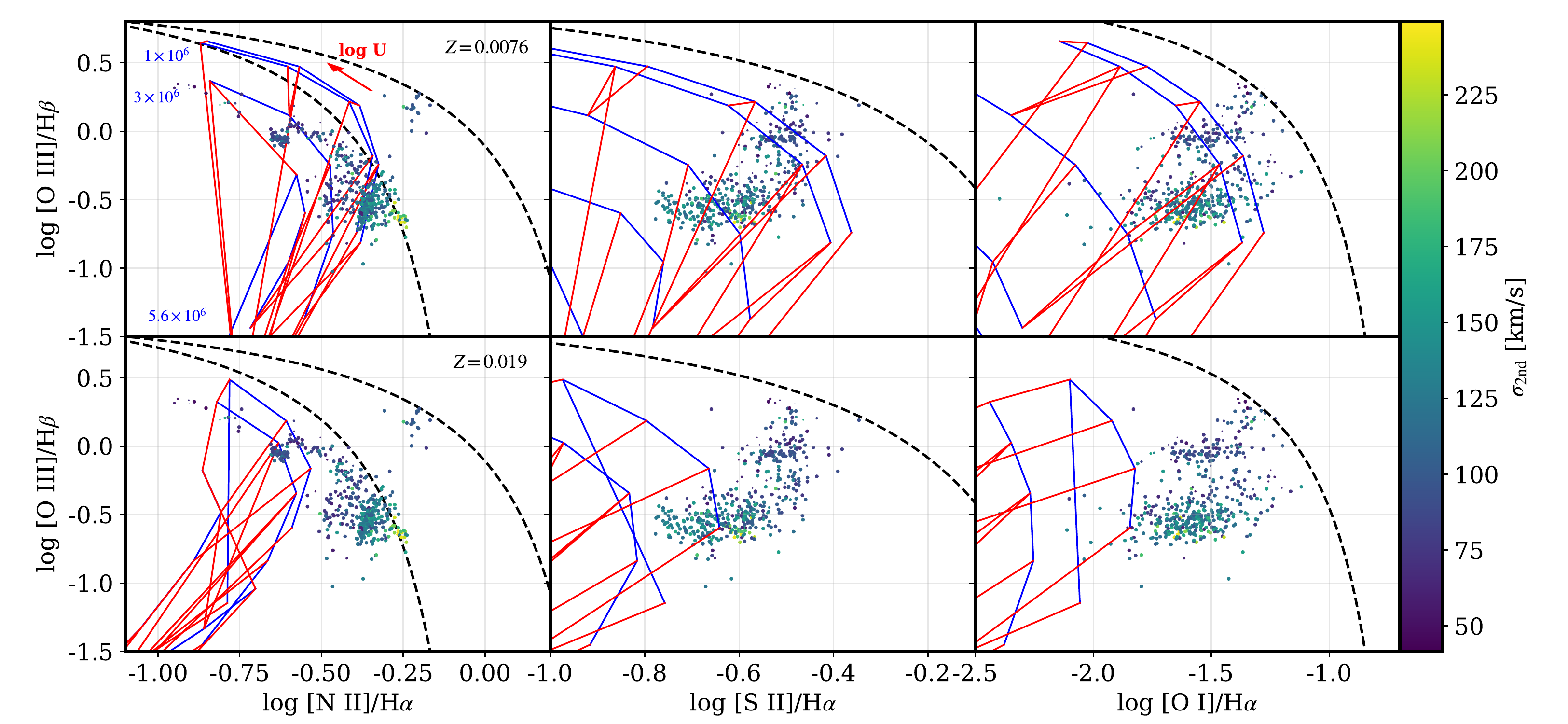}
\caption{BPT diagrams as shown in Fig.~\ref{fig:bpt_diagram}, here in comparison with the photoionization grids obtained from the 3MdB database using \cloudy\,models. The top and bottom panels show the model results for $Z = 0.0076$ and $Z = 0.019$, respectively. For comparison, the sample median metallicity is $Z = 0.0086$. Blue lines represent different stellar ages of $1$, $3$, $4$, $5.6$ and $8.9 \times 10^6$ years as ionization sources, and roughly oriented from top right to bottom left as age increases, although not linearly. Red lines represent variation in the photoionization parameter, ranging from log($U$) between $-4.0$ and $-2.0$ (in steps of $0.5$), roughly oriented from top left to bottom right as $U$ decreases. Some stellar age values are displayed in the top-left panel, along with the increasing $U$ orientation.}
\label{fig:bpt_diagram_cloudy}
\end{figure*}

In order to investigate the gas excitation of the outflowing gas, we analyze the line ratios in the Voronoi regions where we measure a second component in the emission-line spectra, using MUSE resolved data. Fig.~\ref{fig:bpt_diagram} shows the distribution of these regions in the BPT diagrams \citep{baldwin81} and their dependence on the second-component velocity dispersion $\sigma_{\mathrm{2nd}}$. We also display the relation between the \nii/H$\alpha$, \oiii/H$\beta$ and \sii/H$\alpha$ line ratios with $\sigma_{\mathrm{2nd}}$. It is important to note that since we do not let line ratios vary between the two components in the emission-line measurements, we decide to use the line ratio values from the single-component fit, which are better constrained. We observe that the regions where the outflow seems to be more intense ($\sigma_{\mathrm{2nd}} \gtrsim 200\,$\kms) display low \oiii/H$\beta$ ratios and are close to the composite region in the BPT diagram. These values fall quite far from what should be expected by shock ionization models shown by the line grids in the figure, obtained using the \mappings\,code and the 3MdB database \citep{sutherland17,alarie19}. In fact, the regions with the highest $\sigma_{\mathrm{2nd}}$ appear to be the most distant from the regions covered by the shock models.

The dependence of $\sigma_{\mathrm{2nd}}$ on the \nii/H$\alpha$ ratio would indicate that shocks are driving the more turbulent motions. However, this is not in agreement with the \sii/H$\alpha$, a ratio that is more sensitive to shocks, that shows a decrease in regions displaying higher $\sigma_{\mathrm{2nd}}$, where it should also increase. This is illustrated by the dashed lines, which represent how the shock models vary with shock velocities in relation to these line ratios, for several electron densities. The observed \oiii/H$\beta$ shows a similar behavior in comparison to the shock models (it decreases with increasing shock velocities), but with a considerable shift in line ratio values (observed \oiii/H$\beta$ is $\sim 2$ lower than what expected from the models), and this difference causes the observed values to not populate the parameter space of the shock models.

It is important to note that in this comparison we used simple shock models, without the precursor component. Adding this contribution further increases the difference observed in the \oiii/H$\beta$ ratio. 

We also compared the observed line ratios with photoionization models. We obtained emission-line intensities from the 3MdB database for \cloudy\,models \citep{ferland13}, which is described in detail in \citet{morisset15}. In summary, we used the ``CALIFA'' table for H {\sc ii} regions in the 3MdB database, which consists in a \cloudy\,model using the simple stellar populations templates used by \citet{cid13} in the analysis of CALIFA observations. These templates comprise four bins of metallicity ($Z =$ 0.0037, 0.0076, 0.019 and 0.0315) and 39 different stellar ages (between $10^6$ and $1.4 \times 10^{10}$ years). The templates are then used to compute the SED passed to \cloudy\, to run the photoionization models. Aside from the stellar age and metallicity, four other parameters are considered in the 3MdB database: the nebula geometry (``form factor'' $f_{\mathrm{S}} = 0.03$ and $3.0$ for a full sphere and a hollow spherical bubble, respectively); the nitrogen over oxygen abundance ratio (log N/O$ = -0.5$, $-0.25$, $0$, $0.25$ and $0.5$); the ionization parameter ($\mathrm{log(}U$) from $-4$ to $-1.5$ in steps of $0.25$); and the H$\beta$ luminosity fraction of the nebula (H$\beta$ fraction between $\sim20\%$ to $\sim100\%$ in steps of $\sim20\%$).

For this comparison we have used a full sphere geometry ($f_{\mathrm{S}} = 0.03$), a total nebula H$\beta$ luminosity (H$\beta$ fraction $ \sim100\%$) and a log N/O = 0.25. Grids were created using the values of log($U$) between $-4.0$ and $-2.0$ (in steps of $0.5$) and stellar ages of $1$, $3$, $4$, $5.6$ and $8.9 \times 10^6$ years. Fig.~\ref{fig:bpt_diagram_cloudy} shows the BPT diagrams with the data values showed in Fig.~\ref{fig:bpt_diagram} now compared with the photoionization models described above, with metallicity values of $Z = 0.0076$ (upper panels) and $Z = 0.019$ (bottom panels).

The observed line ratios match more closely the grid models for the $Z = 0.0076$ than the ones with $Z = 0.019$. This agrees with the measured median metallicity of the sample, $Z = 0.0086$, being closer to the lower-metallicity models. In general, the observed line ratios are better reproduced by the photionization models than the shock models displayed in Fig.~\ref{fig:bpt_diagram}. The regions presenting high $\sigma_{\mathrm{2nd}}$, observed in the bottom-right corner of the H {\sc ii} classification in the \nii/H$\alpha$ BPT diagram, are closely approached by the photoionization from very young ages ($\sim 1\,$Myr) and low photoionization parameters (log($U$)$\,\sim -4.0$). Slightly older populations ($\sim 5\,$Myr) with also slightly higher photoionization parameters (log($U$)$\,\sim -3.5$) appear to better reproduce the outflow ratios observed in the \sii/H$\alpha$ and \oi/H$\alpha$ BPT diagrams, which present more unfolded model grids and thus could better differentiate the photoionization parameters.

It is worth noting that stellar ages older than $5\,$Myr do not produce a strong \oiii\, emission line due to its high ionization energy level, thus the grids fall to very low \oiii/H$\beta$ ratios (below the limits shown here). \oiii/H$\beta$ would increase again for older populations ($t > 0.5\,$Gyr), due to hot evolved stars, which are not representative of our sample and thus not shown in Fig.~\ref{fig:bpt_diagram_cloudy}.

To summarize, we conclude that the observed line ratios related to the outflowing regions (regions where we measure an emission-line second component) are better represented by photoionization models than compared to shock models, with a sub-solar metallicity. We will further discuss how this compare to other results from the literature in Sec. \ref{sec:disc_ion}.


\section{Outflow parameters}
\label{sec:out}

\begin{figure*}
\centering
\includegraphics[width=\textwidth]{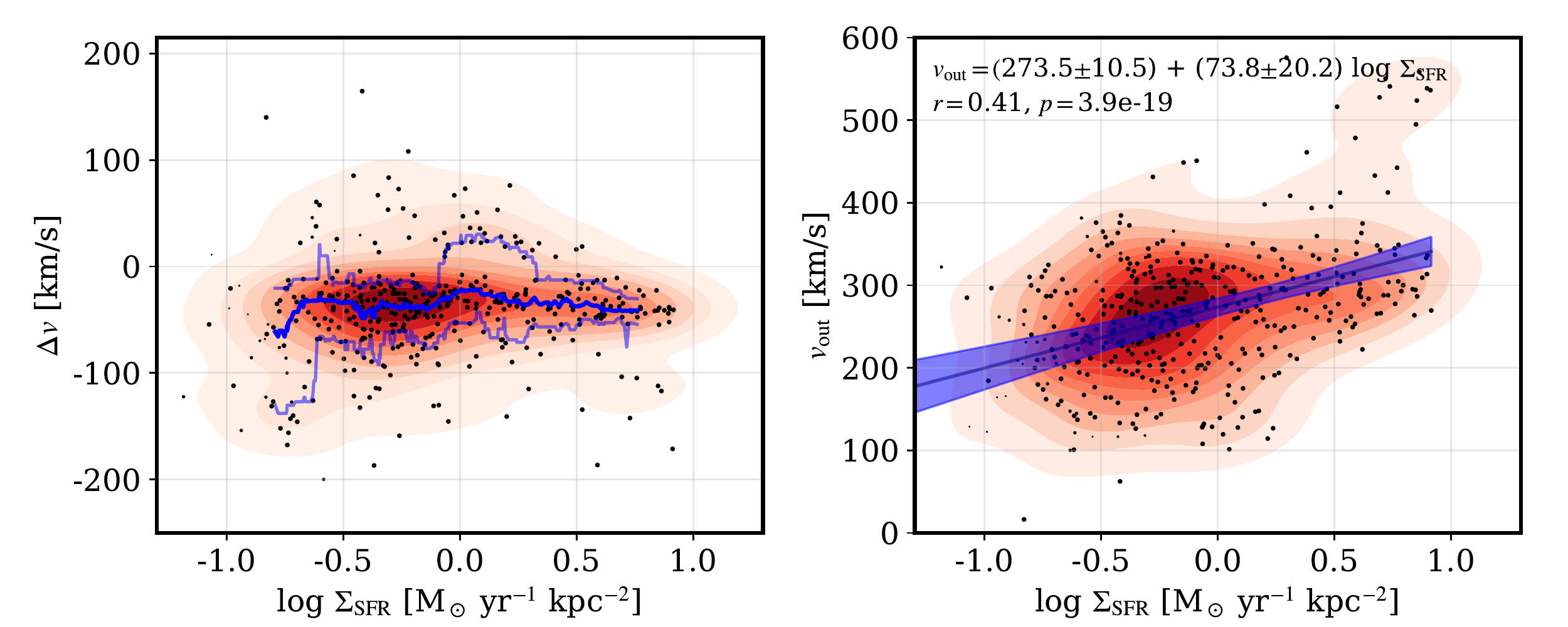}
\caption{Velocity shift between the two emission-line components and outflow velocity as a function of SFR surface density for the Voronoi regions, shown as black points. Blue lines in the left panel display the running median with a 1$\sigma$ distribution. The black line and blue shaded region in the right panel represent the best-fit relation (displayed in the top-left corner, along with Pearson coefficients) with a 3$\sigma$ uncertainty.}
\label{fig:v_out}
\end{figure*}

The second component fitted to the emission-line profiles was detected in 13 of the 15 galaxies in our sample, as displayed in the individual maps for each galaxy in Appendix \ref{app:gal_maps}. Assuming that this second component is related to outflowing gas in the star-forming regions in our galaxy sample, we can use this component to estimate the outflow parameters.
    
\subsection{Outflow velocity}

We first derive the velocity shift between the two components and the outflow velocity, considering also the velocity dispersion of the second component. The velocity shift $\Delta v = v_{\mathrm{1st}} - v_{\mathrm{2nd}}$ as a function of the SFR surface density is shown in the left panel of Fig.~\ref{fig:v_out}. Although we allowed for the second component to assume both positive and negative velocities in relation to the first component, we observe that the outflow velocities are mainly toward blueshifted values, which is expected, since the outflow redshifted component is generally most susceptible to the host-galaxy extinction. The values are concentrated between 0 and $-100\,$\kms, with a large scatter in the lower $\Sigma_{\mathrm{SFR}}$ values, reaching $\sim-40\,$\kms around $\Sigma_{\mathrm{SFR}} \sim 1.0\,$\myrkpc, where the distribution concentrates, and remaining almost constant toward higher $\Sigma_{\mathrm{SFR}}$ values.

We adopt the \citet{genzel11} definition to estimate the maximum outflow velocity, with $v_{\mathrm{out}} \sim | \Delta v - 2\sigma_{\mathrm{2nd}}|$. The right panel of Fig.~\ref{fig:v_out} shows its dependence on $\Sigma_{\mathrm{SFR}}$, where a correlation can be observed. Outflow velocities are usually low, mainly ranging between $200$ to $300\,$\kms, with $v_{\mathrm{out}} \sim 250$\,\kms\, around $\Sigma_{\mathrm{SFR}} \sim 1.0\,$\myrkpc, the peak of the distribution. A branch of higher outflow velocities ($400 < v_{\mathrm{out}} < 550\,$\kms) appears in the higher $\Sigma_{\mathrm{SFR}}$ end, indicating the presence of stronger outflows in the sample, but this region is weakly populated. With a 3$\sigma$ uncertainty, the estimated correlation is described as

\begin{equation}
v_{\mathrm{out}} = (273.5 \pm 10.5) + (73.8 \pm 20.2)\, \mathrm{log} \left( \frac{\Sigma_{\mathrm{SFR}}}{\mathrm{M_\odot\,yr^{-1}\,kpc^{-2}}} \right) .
\end{equation} 

\subsection{Outflow extent and electron density}
\label{sec:out_rad_den}

\begin{figure*}
\centering
\includegraphics[width=\textwidth]{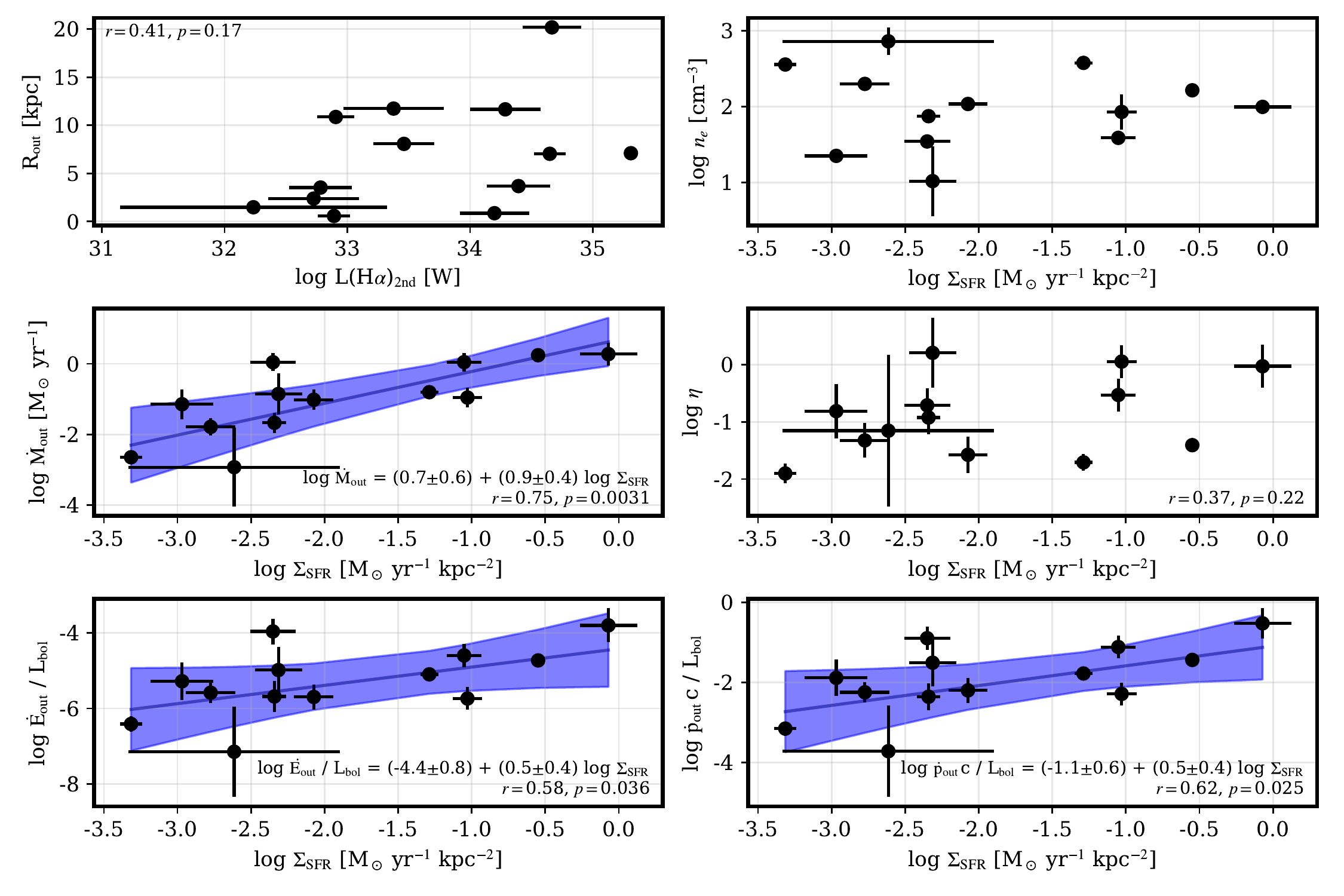}
\caption{Outflow properties of the analyzed sample. All panels are in function of the outflow SFR surface density, except from the top left. Top left: outflow extent (definition described in the text) as a function of the second-component H$\alpha$ luminosity. Top right: electron density. Middle left: mass outflow rate. Middle right: mass loading factor. Bottom left: ratio between the outflow kinetic power and the total bolometric luminosity. Bottom right: ratio between the momentum outflow rate and the SFR photon momentum. We display the Pearson coefficients for five of the displayed relations, and the best-fit model with a 3$\sigma$ uncertainty for the panels shown as black line with blue-shaded regions.}
\label{fig:out_params}
\end{figure*}

The estimation of the outflow extent, or outflow radius, and the outflow geometry in general, is perhaps one of the parameters leading to the higher uncertainties in outflow characterization. Here we have assumed that we do not spatially resolve individual H {\sc ii} regions\footnote{The median seeing-limited  spatial resolution of 1.1 kpc for our sample (see Sec.~\ref{sec:muse}) is larger than expected for a typical H {\sc ii} region size.}, and thus we also do not resolve the outflow spatially. However, we do detect the emission-line second component in regions smaller than the spatial resolution.

In order to deal with this issue, we have only estimated the total outflow extent within each galaxy in the sample. This was performed by summing up the spaxels area in kpc$^2$, and assuming a projected circular geometry. Thus the resulting outflow projected area can be represented by $A_{\mathrm{out}} = \pi R_{\mathrm{out}}^2$, where $R_{\mathrm{out}}$ is the radius of the entire outflow area in the galaxy. The estimated $R_{\mathrm{out}}$ values are displayed in the top-left panel of Fig.~\ref{fig:out_params}, as well as its relation with the second-component H$\alpha$ luminosity. A weak positive relation is observed with a large scatter. Outflow extents vary between 0.5 and 20 kpc, with an average of 6.8 kpc and a 1$\sigma$ distribution of 5.4 kpc.

We have also estimated the outflow electron density, as it is necessary to further estimate the outflow energetics. The [S\,{\sc ii}] emission lines are quite weak within the Voronoi regions, resulting in unconstrained estimations of the electron density. We thus estimated the electron density by integrating the fluxes of both lines in the regions where we detect the outflowing component, and used the ratio from these integrated fluxes in the calculation. The electron densities were obtained by using the \pyneb\,package \citep{luridiana15} assuming a nebular temperature of $10\,000\,$K, and are shown as a function of the SFR surface density in the top-right panel of Fig.~\ref{fig:out_params}. Density values vary between 10 and 730 \cmt, with a mean value of $166.8 \substack{+29.3 \\ -28.7}$ \cmt.

It is worth noting that the SFR surface density values displayed in Fig.~\ref{fig:out_params} are obtained within the integrated outflowing regions in each galaxy, divided by the outflowing area (calculated as mentioned above). Thus these values should not be directly compared to the SFR surface density values displayed in the figures previously showed, which take into account the SFR within the Voronoi regions.

\subsection{Mass outflow rate and loading factor}

After estimating outflow properties such as velocity, density and extent, we can now derive the mass outflow rate and its relation with the star formation rate. Following the assumption that the outflow does not vary radially and that the second component is indeed related to the outflowing gas (either by the expanding shocked gas or by photoionization), we can derive the mass outflow rate as \citep{newman12}:

\begin{equation}
\dot{M}_{\mathrm{out}} = \frac{1.36\,m_{\mathrm{H}}}{\gamma_{\mathrm{H\alpha}}\,n_e} \left( \frac{v_{\mathrm{out}}}{R_{\mathrm{out}}} \right) \, L(\mathrm{H\alpha})_{\mathrm{2nd}}  ,
\end{equation} 

\noindent
where $m_{\mathrm{H}}$ is the hydrogen atomic mass, $\gamma_{\mathrm{H\alpha}}$ is the H$\alpha$ emissivity at gas temperature of $T_e = 10^4\,$K ($\gamma_{\mathrm{H\alpha}} = 3.56 \times 10^{-25}\,{\mathrm{erg\,cm^{3}\,s^{-1}}}$), $n_e$ is the outflow electron density, $v_{\mathrm{out}}$ and $R_{\mathrm{out}}$ are the outflow velocity and radial extent, respectively, and the $L(\mathrm{H\alpha})_{\mathrm{2nd}}$ is the H$\alpha$ luminosity of the second component, related to the outflowing gas.

We used the $R_{\mathrm{out}}$ and $n_e$ values described in the previous section, and displayed in the top panels of Fig.~\ref{fig:out_params}. For $v_{\mathrm{out}}$, we used a median value for each galaxy and the $L(\mathrm{H\alpha})_{\mathrm{2nd}}$ values used in this calculation are also shown in the top-left panel of Fig.~\ref{fig:out_params}. The relation between the derived mass outflow rate and the SFR surface density is shown in the middle-left panel of Fig.~\ref{fig:out_params}, and displays a strong correlation ($r = 0.75$) with a p-value indicating less than 0.5\% chance of the relation originating at random. Mass outflow rate values vary between $1.2 \times 10^{-3}\,$\myr and $1.9\,$\myr, with a median of $0.1\,$\myr. The best-fit relation obtained is given by

\begin{equation}
\dot{M}_{\mathrm{out}} = (0.7 \pm 0.6) + (0.9 \pm 0.4)\, \mathrm{log} \left( \frac{\Sigma_{\mathrm{SFR}}}{\mathrm{M_\odot\,yr^{-1}\,kpc^{-2}}} \right) ,
\end{equation} 

\noindent
considering a 3$\sigma$ uncertainty.

The middle-right panel of Fig.~\ref{fig:out_params} shows the relation between the SFR surface density and the mass loading factor, given by $\eta = \dot{M}_{\mathrm{out}}/$SFR. Although there seems to be a positive correlation, there is a considerable scatter ($p{\mathrm{-value}} = 0.22$) indicating that a larger galaxy sample would be necessary to confirm this trend. We will further discuss these results in Sec. \ref{sec:out_prop_disc}.

\subsection{Outflow energetics}

Finally we use the derived outflow properties to obtain the outflow kinetic power and momentum rate. The outflow kinetic power is given by $\dot{E}_{\mathrm{out}} = {\frac{1}{2}}\, \dot{M}_{\mathrm{out}}\, v_{\mathrm{out}}^2$, and its dependence with $\Sigma\,$SFR is shown in the bottom-left panel of Fig.~\ref{fig:out_params}, displayed as the ratio with the bolometric luminosity assuming that $L_{\mathrm{bol}} \sim 10^{10}\times$SFR$\,L_\odot$ \citep{kennicutt98}. The outflow momentum rate is given by  $\dot{p}_{\mathrm{out}} = \dot{M}_{\mathrm{out}}\, v_{\mathrm{out}}^2$, and is displayed in the bottom-right panel of Fig.~\ref{fig:out_params} as a ratio to the photon momentum rate derived from the bolometric luminosity.

As expected from the relation between the $\dot{M}_{\mathrm{out}}$ and $\Sigma_{\mathrm{SFR}}$ both $\dot{E}_{\mathrm{out}}$ and $\dot{p}_{\mathrm{out}}$ also relate with SFR surface density, although with a weaker correlation and an increased scatter. The best-fit relations are

\begin{equation}
\mathrm{log}\,\frac{\dot{E}_{\mathrm{out}}}{L_{\mathrm{bol}}} = (-4.4 \pm 0.8) + (0.5 \pm 0.4)\, \mathrm{log} \left( \frac{\Sigma_{\mathrm{SFR}}}{\mathrm{M_\odot\,yr^{-1}\,kpc^{-2}}} \right),
\end{equation} 

\begin{equation}
\mathrm{log}\,\frac{\dot{p}_{\mathrm{out}}}{L_{\mathrm{bol}}/c} = (-1.1 \pm 0.6) + (0.5 \pm 0.4)\, \mathrm{log} \left( \frac{\Sigma_{\mathrm{SFR}}}{\mathrm{M_\odot\,yr^{-1}\,kpc^{-2}}} \right),
\end{equation} 

\noindent
where $c$ is the light speed.

Outflow kinetic power accounts for very little, with values varying between $7.2 \times 10^{-6}\%$ and $0.02\%$ of the bolometric luminosity, with a median value of only $5.2 \times 10^{-4}\%$. The outflow, however, represents a median value of 1.3\% of the momentum driven by the bolometric luminosity alone (varying between 0.02\% and 30\%), indicating that although the outflow does not contribute much to the energy deposition in the ISM, its contribution is not as negligible to the gas kinematics. 


\section{Discussion}
\label{sec:disc}

\subsection{Feedback impact on SFR}

\begin{figure}
\centering
\includegraphics[width=\columnwidth]{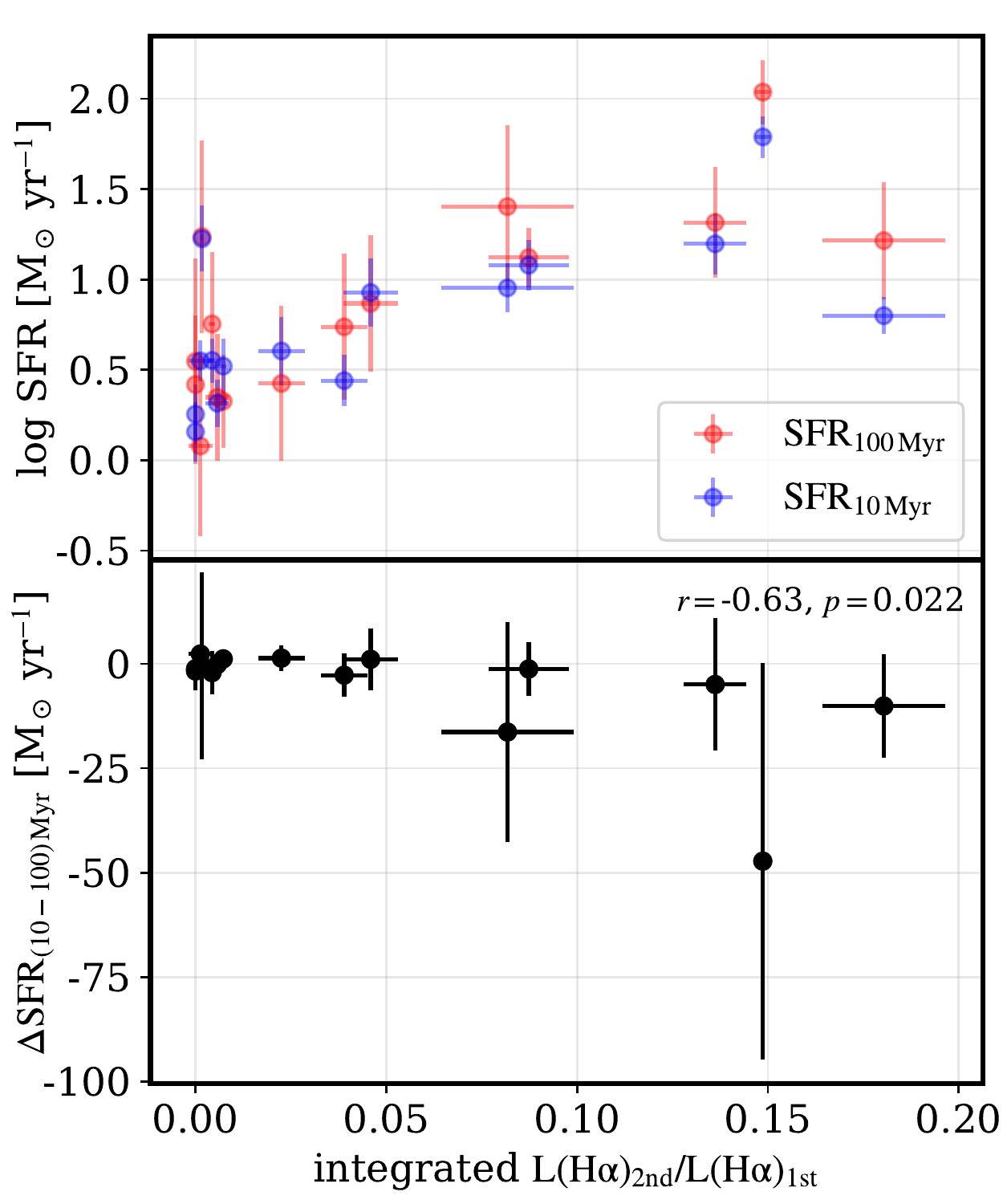}
\caption{Relation between the H$\alpha$ luminosity ratio between the second and first components and the SFR measured at 10 (blue) and 100 Myr (red points, top panel) and the difference of SFR between these two epochs (bottom panel). Points at the same x-axis value in the top panel represent the same galaxy at different timescales. The Pearson coefficients in the bottom panel indicate a moderate anticorrelation with a higher significance than observed in previous comparisons made in Figs.~\ref{fig:sfr_ratio_slope} and \ref{fig:lum_ratio_vs_sfr_ratio}.}
\label{fig:lha_sfr}
\end{figure}

The results obtained with the SED fitting reported in Sec. \ref{sec:sed} indicate no
clear feedback impact in the recent SFH probed by our sample. We observe in Figs. \ref{fig:sfr_ratio_slope} and \ref{fig:lum_ratio_vs_sfr_ratio} that the ratio between the SFRs at 10 Myr and at 100 Myr in entire galaxies does not change significantly either with single- or with double-component properties. These results indicate that even though we can trace outflowing gas signatures, the feedback is not strong enough to impact the recent star-forming activity.

We explore these results further by investigating not only the relation between the H$\alpha$ luminosity ratio and the SFR ratio, but with SFRs at 10 and 100 Myrs and their difference, displayed in Fig.~\ref{fig:lha_sfr}. We can observe that the outflow has increased luminosity contribution in higher-SFR galaxies in both timescales. The bottom panel of Fig.~\ref{fig:lha_sfr} also indicates a decrease of SFR between 100 and 10 Myr (up to $\sim 100$ \myr) for some galaxies presenting higher outflow contribution. This relation exhibits a much higher significance ($p$-value = 0.02) when compared to the other relations showed in Figs. \ref{fig:sfr_ratio_slope} and \ref{fig:lum_ratio_vs_sfr_ratio}, and follows the trend suggested in Fig.~\ref{fig:lum_ratio_vs_sfr_ratio}, where the SFR at 10 Myr is decreased in relation to the SFR at 100 Myr. However, as indicated in Fig.~\ref{fig:lha_sfr} top panel, the SFR decrease at 10 Myr is minor, not enough to change the overall trend of increasing SFR with increasing outflow luminosity.

To summarize, although we can observe outflow H$\alpha$ luminosity contributions of up to $\sim 20\%$ of the rotating ionized gas, observed in the higher-SFR galaxies, we only observe a clear change in recent SFR when comparing absolute values between 10 and 100 Myr. Circumstantial evidences, which could be confirmed with a larger sample, indicate in the same direction when comparing the outflow luminosity contribution with the SFR ratio, and when looking into the individual slopes between the velocity residuals and the velocity dispersion. These results point towards a scenario where the feedback is responsible for suppressing some of the star formation between 100 and 10 Myr timescales.

\subsection{Outflow ionization mechanism}
\label{sec:disc_ion}

As discussed in Sec. \ref{sec:gas_exc}, the comparison of the observed line ratios favors a scenario where the outflowing gas is driven by photoionization, while shock models can not reproduce the data distribution in the BPT diagrams. These results are somewhat surprising, because outflows in star-forming galaxies have been linked to shock excitation in the literature. \citet{rodriguezdelpino19} have reported, in theirs study of nearby ($z < 0.015$) MaNGA star-forming galaxies, a correlation between the \sii/H$\alpha$ line ratio, a known tracer of shocks, and the velocity dispersion of both narrow and broad emission line components, which is not observed in our data (see Fig.~\ref{fig:bpt_diagram}). \citet{lopezcoba17}, comparing the outflows in the inclined disk galaxy UGC 10043 ($z \sim 0.007$, $\mathrm{SFR} = 0.36\,$\myr, log($\mathrm{M_*/M_\odot}$) = 9.79) to the \mappingsiii\,photoionization and shock models, in a very similar fashion as showed in this work, have concluded that shocks indeed better describe the observed line ratios. Other studies relating star-formation-driven local universe feedback showing signatures of strong shocks include \citet{sharp10}, \citet{ho14} and \citet{avery21}, among others.

Although shock-induced feedback seems to be a majority in the reports found in the literature, some do find evidence that photoionization is the main outflow ionization mechanism in star-forming galaxies. \citet{davies19} have reported low \nii/H$\alpha$ ratio values for both narrow and broad components in $z \sim 2.3\,$ star-forming sample ($9.4 <$ log($\mathrm{M_*/M_\odot}$) $< 10.8$, median log($\mathrm{M_*/M_\odot}$) = 10.1; main-sequence offset of $-0.8 <$ log($\mathrm{SFR/SFR_{MS}}$) $< 1.0$, median log($\mathrm{SFR/SFR_{MS}}$) = 0.3), indicating low contribution of shocks. They also noted the results from \citet{sharp10}, which can give some helpful insights to our results. Studying a sample of 5 nearby starburst galaxies ($z < 0.03$), \citet{sharp10} modeled the time evolution of ionizing and mechanical luminosities that arise from star-formation activity, mainly separating the photoionization and the wind-driven shocks mechanisms, for an instantaneous and a continuous star-forming scenarios (see their Fig. 29). For the starburst scenario the photoionization from young stars dominates up to $\sim 10\,$Myr, when then shocks from winds created by the radiation pressure increase in luminosity contribution, dominating the luminosity emission up until $\sim 100\,$Myr, when both mechanisms fall to negligible values. On the other hand, the continuous star-formation scenario indicates a domination of the photoionization mechanism over wind-driven shocks in all time evolution in general.

As we observe that the photoionization mechanism dominates, the scenario proposed by \citet{sharp10} is in agreement either with the early stages of the starburst model or with any timescale of the continuous star-formation model. As mentioned in the previous section, we do not see any major changes between the SFRs at 10 and 100 Myr (see Fig.~\ref{fig:lha_sfr}), indicating that the galaxies in our sample are better reproduced by the continuous star-formation scenario. Indeed, since these are galaxies lying close to the star-forming main sequence, this seems to be a reasonable scenario. In this case, the \citet{sharp10} results do not help us to further constrain the outflow timescale, as the photoionization model dominates at all times.

\subsection{Outflow properties}
\label{sec:out_prop_disc}

As further studies are performed in star-forming feedback, better we constrain the gas outflow properties. However, some properties are harder to constrain than others, which sometimes drives differences between studies when used differently. These are specially the case of the outflow electron density and radial extent. In this section we try to compare these properties with other results in the literature.

As discussed in Sec. \ref{sec:out_rad_den}, the outflow extent used in our work is not conventional, since we integrate the entire region where we find outflowing gas for each galaxy. The outflow radius for each galaxy is displayed in Fig.~\ref{fig:out_params} and it has a mean value of $R_{\mathrm{out}} = 6.8\,$kpc. While studying a large sample of low-redshift ($z < 0.2$) luminous infrared galaxies (LIRGS), \citet{arribas14} reported that outflows are generally quite compact, usually $\sim 0.7\,$kpc in radius, although it can reach up to $\sim 2\,$kpc for some galaxies. \citet{davies19} used the mean value between the PSF-resolved spatial resolution ($0.7\,$kpc) and maximum radius extent observed in their sample ($2.6\,$kpc), $R_{\mathrm{out}} = 1.7\,$kpc. Assuming biconical geometry outflows in a MaNGA galaxy subsample, \citet{bizyaev19} derived that the cone radius is typically $\sim 1\,$kpc in the disk plane. These are all smaller values than we have used, which is expected given the different methodologies.

Regarding the outflow electron density, changes should not be as dramatic, as it is an intensive property, in principle characteristic of the outflowing gas regardless of the area it covers. We find density values between 10 and 730 \cmt, with a mean value of $166.8 \substack{+29.3 \\ -28.7}$ \cmt, while the papers mentioned above find (or use) $70\,$\cmt\, \citep{bizyaev19}, $380 \substack{+249 \\ -167}$ \cmt\, \citep{davies19} and $315\,$\cmt\, \citep{arribas14}. Higher values from \citet{arribas14} and \citet{davies19} are probably due to the higher-SFR nature of their galaxy sample, while \citet{bizyaev19} electron density could be affected by their outflow geometry assumption, lowering its value in comparison to ours. Regardless of the origins of these differences, they are not large.

\begin{figure}
\centering
\includegraphics[width=\columnwidth]{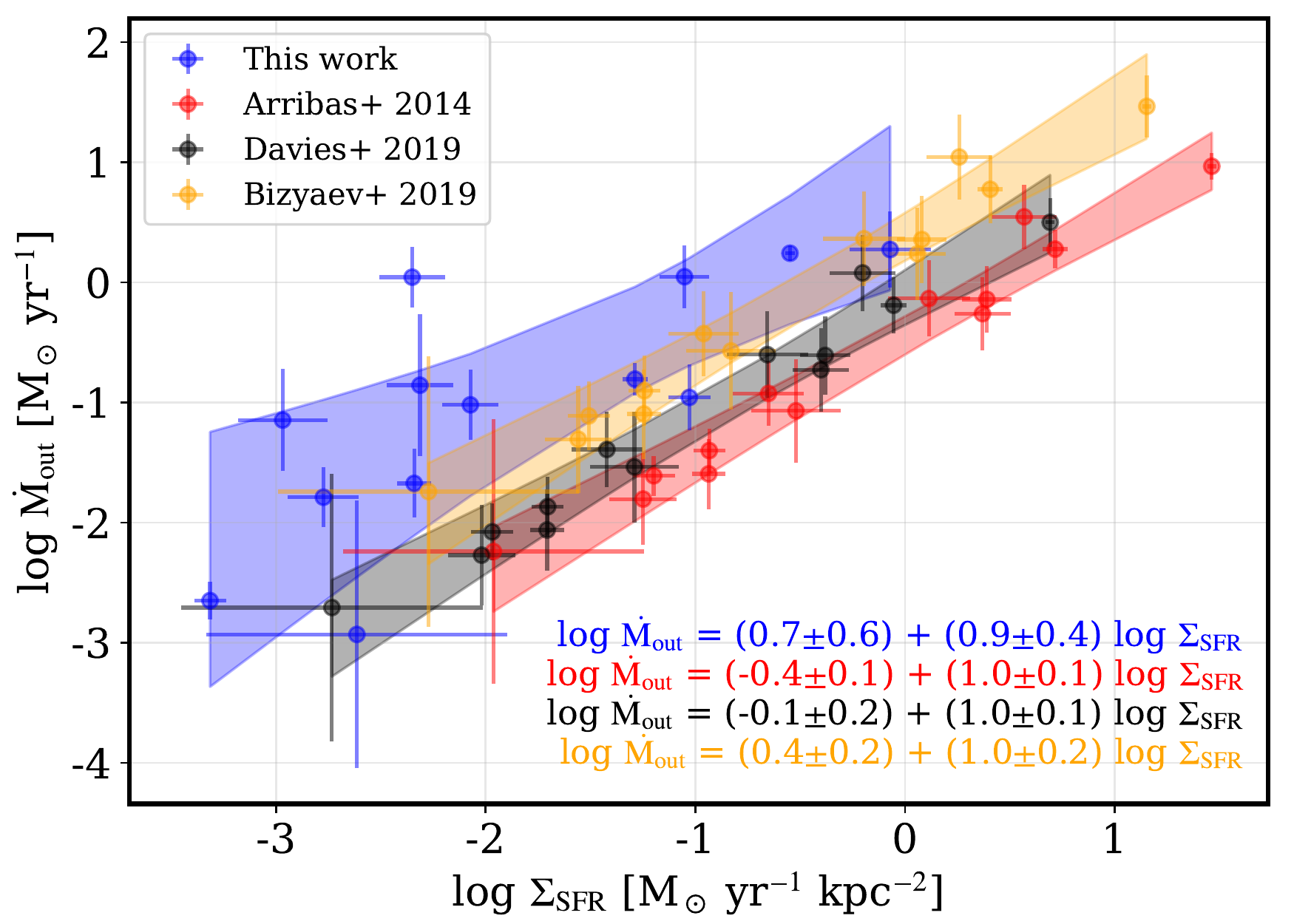}
\caption{Mass outflow rate as a function of the SFR surface density using outflow radial extent and electron density values from different articles. As in Fig.~\ref{fig:out_params}, shaded regions display the 3$\sigma$ distributions, and the equations in the bottom-right corner describe the best-fit models.}
\label{fig:out_params_diff}
\end{figure}

Fig.~\ref{fig:out_params_diff} displays the difference in the mass outflow rate dependence with the SFR surface density using the values mentioned for these studies. The curves present roughly the same inclination, indicating that the mass loading factor does not vary with these different property measurements, aside from a constant represented by the shift in mass outflow rate for a given SFR surface density. With this comparison we conclude that regardless of the radial extent and electron density measurement strategy we apply, the main results regarding outflow scaling relations are not affected materially.

\subsection{Comparison with other observational results from the literature}
\label{sec:obslit_comp}

The results we have obtained for the outflow properties are different than other studies regarding the mass outflow rate and the mass loading factor, usually indicating that we are tracing weaker outflows, but maybe expected for our galaxy sample, which presents moderate star-formation activity.

The slope between the mass outflow rate and $\Sigma_{\mathrm{SFR}}$ we obtain in this work ($0.9\pm0.3$) is considerably steeper than the reported in MaNGA outflows in nearby galaxies by \citet{avery21} ($0.26\pm0.08$). In their case, they do not separate the AGN contribution in this relation, which could account for this difference since AGN emission is higher in the central parts that usually contribute less to $\Sigma_{\mathrm{SFR}}$ than disk regions. Those authors found no correlation between the $\Sigma_{\mathrm{SFR}}$ and $\eta$, not far from our results, where we find a quite weak positive correlation, but with a very scattered distribution.

The traced outflows in our sample are in general quite weak in comparison with other studies, not only represented by the very low contribution to the bolometric luminosity ($\sim 5.2 \times 10^{-4}\%$), but also in the mass loading factor, with values of $0.01 < \eta < 1.6$ with a median of $0.12$. This is lower than reported by \citet{rodriguezdelpino19} for their star-forming MaNGA sample (median value $\eta \sim 0.25$) and more distant to the values reported for galaxies with higher SFRs \citep[$\eta \sim$ 0.3--0.5,][]{arribas14,davies19}. \citet{avery21} report values of outflow momentum rate $\sim 10\%$ of total momentum output, including AGN (although the AGN luminosity contribution is usually less than 20\%), while we observe $\sim 0.7\%$, even though their $\eta$ values seem to be similar to what we obtain (see their Fig. 11). In another study using MaNGA data, tracing outflows using NaD neutral gas, \citet{robertsborsani20} obtain $\eta$ values also somewhat higher than we do ($-1 < \mathrm{log}\, \eta < 1$, with a median probably $\eta \gtrsim 1$). \citet{zaragozacardiel20}, using a star-formation self-regulator model in a large sample of galaxy discs with MUSE data, find $1 < \eta < 5$ for $7.5 < \mathrm{log}\, \Sigma_* < 9\,\mathrm{M_\odot\,kpc^{-2}}$.

As discussed in Sec. \ref{sec:out_prop_disc}, the outflow properties can change the estimated contribution, also affecting the mass loading factor. If we use the outflow extent and the electron density from \citet{arribas14} and \citet{davies19}, however, we obtain median values of $\eta = 0.22$ and $\eta = 0.08$, respectively, still lower than the reported values. This indicates that the luminosity contribution of the outflows is indeed quite low for the SFR regime probed by our sample. Thus the discrepancy between our results and the others from the literature could be explained by either the deeper observations (in comparison to MaNGA data, for example) or the lower SFRs (e.g., in comparison to U/LIRGS).

\subsection{Comparison with theoretical results}
\label{sec:modlit_comp}

Although our mass loading factor values do not exactly match the values reported from other observational studies, it seems to be in agreement with theoretical studies. FIRE simulations presented by \citet{muratov15}, modelling small-scale feedback up to $z = 0$, indicate that mass loading factors for disc progenitors decrease to $\eta \ll 1$ at low redshift. Using a small-box simulation resolving supernova remnants in the ISM, \citet{li20} report values of $\eta \gtrsim 0.08$ for their breakout wind ISM condition, where hot outflows are considered, higher than the partial breakout and cold outflow models ($0.01 < \eta < 0.08$ and $\eta < 0.01$, respectively). \citet{schneider20} also report a $0.1 < \eta < 0.2$ for the outflow hot gas phase, depending on the outflow radial extent. 

\begin{figure}
\centering
\includegraphics[width=\columnwidth]{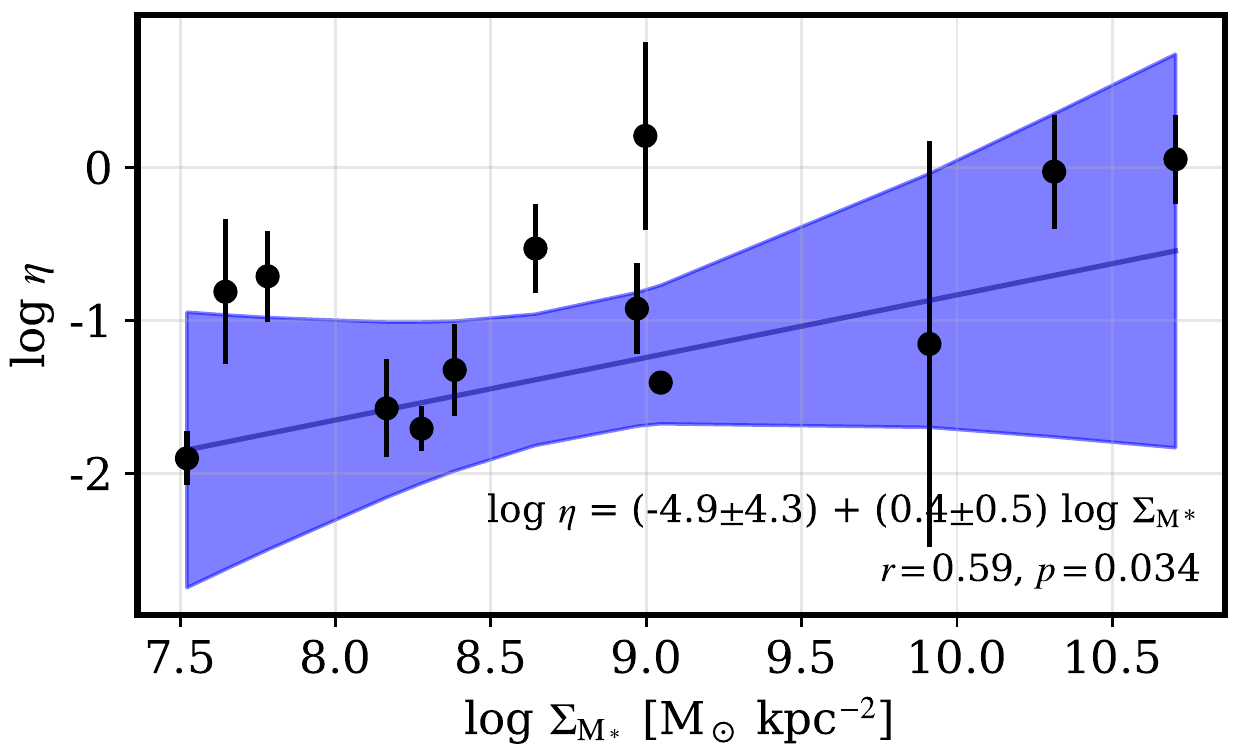}
\caption{Relation between the stellar-mass surface density and the mass loading factor. As in the other figures, the best-fit model with a 3$\sigma$ uncertainty is represented by the black and the blue-shaded region, and described by the equation along with the Pearson coefficients.}
\label{fig:out_params_mass}
\end{figure}

Usually feedback theoretical studies prefer to characterize the outflow properties, such as the mass loading factor, along with the stellar mass instead of the SFR, as we have used throughout this work. Fig.~\ref{fig:out_params_mass} displays the dependence of the mass loading factor with the stellar-mass surface density. $\eta$ values are the same as displayed in Fig.~\ref{fig:out_params}, while $\Sigma_\mathrm{M_*}$ was obtained integrating the stellar mass derived from \starlight\, fit within the same outflowing radius determined in Sec.~\ref{sec:out_rad_den}. We observe a clearer correlation between these parameters, in contrast with the comparison between $\eta$ and $\Sigma_\mathrm{SFR}$. The obtained relation is given by

\begin{equation}
\mathrm{log}\, \eta = (-4.9 \pm 4.3) + (0.4 \pm 0.5)\, \mathrm{log} \left( \frac{\Sigma_{\mathrm{M_*}}}{\mathrm{M_\odot\,kpc^{-2}}} \right).
\end{equation} 

The correlation between the mass loading factor and the SFR or stellar mass has been debated in recent studies. Although observational relations have been observed to be positive in several scenarios \citep[e.g.][including this work; although \citet{zaragozacardiel20} do find an anticorrelation]{arribas14,davies19, avery21}, models seem to indicate otherwise \citep[e.g.][]{creasey13,muratov15,li17}. Several aspects of the feedback characterization could be driving this discrepancy, such as the different gas phase measurements, physical properties (as discussed in Sec.~\ref{sec:out_prop_disc}), and the galaxy scales probed. Further constraints in this topic seem to be necessary to understand what is the reason for this slope uncertainty.


\section{Conclusions}
\label{sec:conc}

In this work we have used detailed MUSE and GAMA observational data of 15 low-redshift ($z \sim 0.15$) star-forming galaxies included in the VALES sample in order to characterize the ionized gas feedback due to star-formation activity. 

We initially measured the emission-line spectra with a single component, in order to investigate the kinematical deviation due to feedback from the galaxy rotation, which we have modelled. We do not observe evidence of increase in residual velocities $v_{\mathrm{res}}$ when compared to the velocity dispersion $\sigma$ and SFR surface density $\Sigma_{\mathrm{SFR}}$ for the entire sample, although we observe for some individual galaxies. We then applied a double-component fit for galaxy regions where the initial measurements were not satisfactory, in order to trace a broader component, interpreted to be related with outflowing gas. The positive relation between the second-component velocity dispersion $\sigma_{\mathrm{2nd}}$ and $\Sigma_{\mathrm{SFR}}$ strengthens this interpretation. 

We used the results obtained from the ionized gas emission lines to compare with the SED modelling of the GAMA photometric broadbands. Modelling the star-formation histories of the sample we could conclude that the SFR at 10 Myr is somewhat suppressed in galaxies with higher H$\alpha$ luminosity contribution from outflows, in comparison with the SFR at 100 Myr ago. The effect of the feedback in these timescales is also observed in comparison to the $\sigma$ versus $v_{\mathrm{res}}$ slope for individual galaxies obtained from the single-component emission-line fit, although with a less significant statistics.

When comparing the gas excitation with shocks and photoionization models, we see that the observed line ratios are better reproduced by the ionization from young stellar population radiation, in contrast with the bulk of recent reports on feedback analysis, which suggest that shocks are most common in stellar feedback. This difference could be explained by the broadly constant recent star formation ($< 100$\,Myr), while a more bursty activity could result in shock-dominated ionization.

The derived outflow properties, such as mass outflow rate ($\sim 0.1\,$\myr), outflow kinetic power ($\sim 5.2 \times 10^{-4}\% L_{\mathrm{bol}}$) and mass loading factor ($\sim 0.12$) indicate that the outflows we observe are weak, and have probably little effect on the galaxies' evolution. Even though stellar feedback seems to be a common phenomenon within star-forming galaxies, as we have detected outflowing regions in 13 out of 15 galaxies, our results indicate that feedback in low-redshift star-forming galaxies is not effective in large gas removal and dramatic SF quenching, resulting in minor (and apparently local) effects within these galaxies.


\section*{Acknowledgments}

GSC acknowledges the support from the Comité Mixto ESO-Chile, the DGI at University of Antofagasta, and from CONICYT/ANID FONDECYT project No. 3190561. GSC would also like to thank Laurent Chemin for insightful discussions about the rotation model results in this work. TMH acknowledges the support from the Chinese Academy of Science (CAS) and the National Commission for Scientific and Technological Research of Chile (CONICYT) through a CAS-CONICYT Joint Postdoctoral Fellowship administered by the CAS South America Center for Astronomy (CASSACA) in Santiago, Chile. MB acknowledges partial support from FONDECYT regular grant 1170618 and 1211000. EI acknowledges partial support from FONDECYT through grant N$^\circ$\,1171710. R.L. acknowledges support from CATA, BASAL grant AFB-170002. YQX acknowledges support from NSFC-12025303, 11890693, 11421303, the CAS Frontier Science Key Research Program (QYZDJ-SSW-SLH006), the K.C. Wong Education Foundation, and the science research grants from the China Manned Space Project with NO. CMS-CSST-2021-A06.


\bibliographystyle{aa}
\bibliography{refs}


\appendix

\section{Parameters of individual galaxies}
\label{app:gal_maps}

\begin{figure*}
\centering
\includegraphics[width=\textwidth]{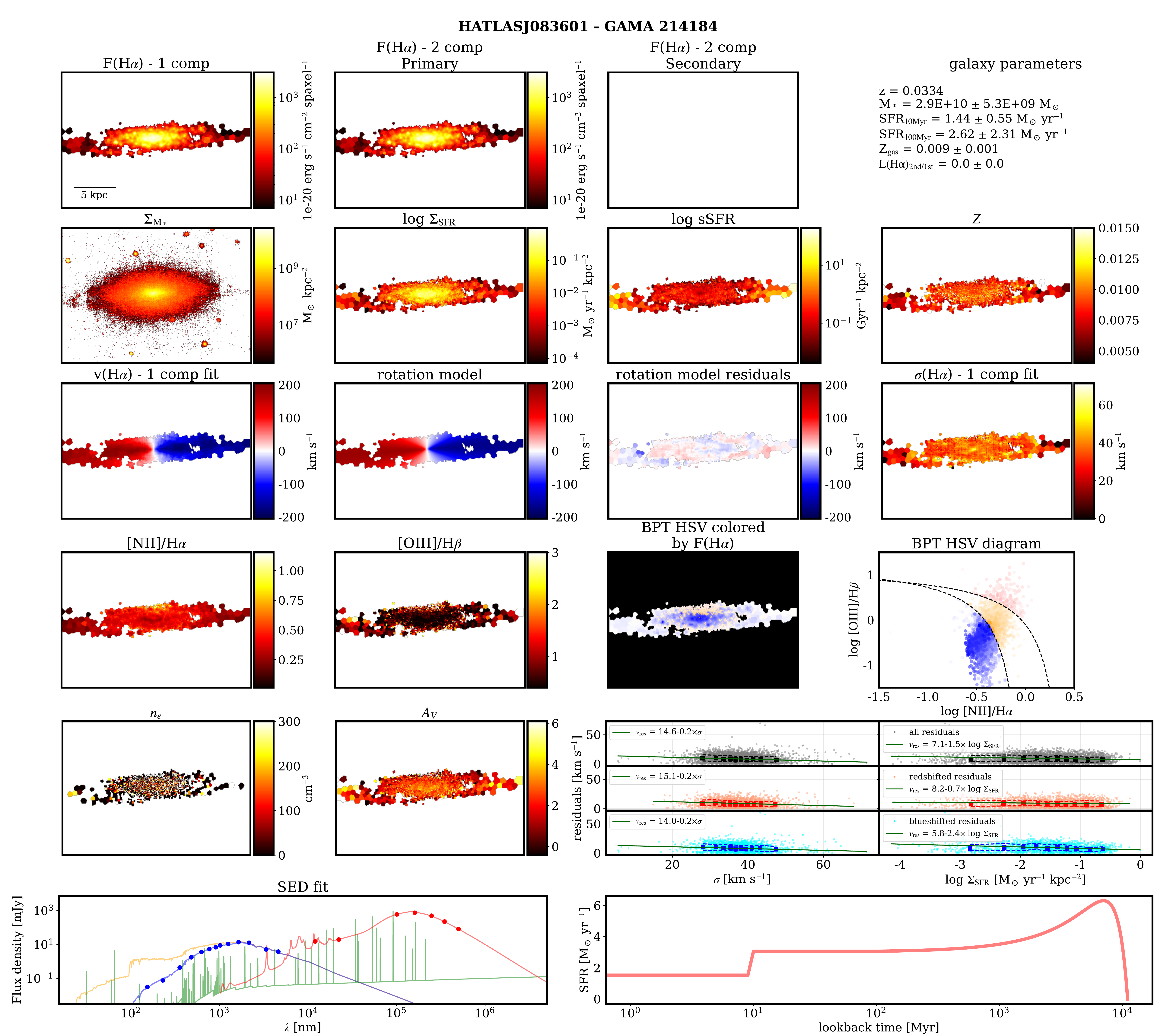}
\caption{Individual data of HATLASJ083601. Listed from top left to bottom right: (first row) H$\alpha$ emission-line integrated fluxes of the one-component fit and the first and second component of the two component-fit maps; (second row) maps of the stellar-mass surface density, SFR surface density in logarithmic scale, sSFR surface density in logarithmic scale, nebular metallicity; (third row) maps of the H$\alpha$ peak velocity, rotation model, residuals between the latter two and velocity dispersion; (fourth row) maps of the \nii/H$\alpha$ and \oiii/H$\beta$ ratios, BPT coloring weighted by the H$\alpha$ flux (using a HSV (hue, saturation, value) color scheme) and BPT diagram; (fifth row) maps of the electron density and optical extinction, and the relations between velocity dispersion and SFR surface density with the rotation model residual velocities; (sixth row) SED fit from \cigale\, and star-formation histories obtained from this fit. A list of physical parameters is also listed in the top-right corner of the figure. Aside from the identified H$\alpha$ flux maps, all maps are retrieved from emission-line measurements. Stellar masses are obtained from \starlight\,stellar continuum fit. Nebular metallicity was estimated using the O3N2 calibrator with the relation presented by \citet{marino13}. The BPT diagram \citep{baldwin81} also displays in dashed lines the commonly used theoretical \citep{kewley01} and empirical \citep{kauffmann03} lines which aim to separate pure star formation and star formation combined with AGN ionization, respectively. Each point in the BPT diagram represents a Voronoi region. Electron densities are obtained by using the \pyneb\,package \citep{luridiana15} and the \sii$\lambda6717/6731$ ratio, assuming a nebular temperature of $10\,000\,$K. Plots of the dependences of residual velocities with velocity dispersion and SFR surface density are separated into all Voronoi regions, just regions presenting redshifted residuals and regions presenting blueshifted residuals, from top to bottom, respectively. Green lines represent a simple fit between the parameters. Finally, the SED fit shows the UV/optical and infrared bands fitted in blue and red dots. Blue, orange, green and red lines represent stellar attenuated and unattenuated, nebular and dust emission, respectively.}
\label{fig:HATLASJ083601}
\end{figure*}

\begin{figure*}
\centering
\includegraphics[width=\textwidth]{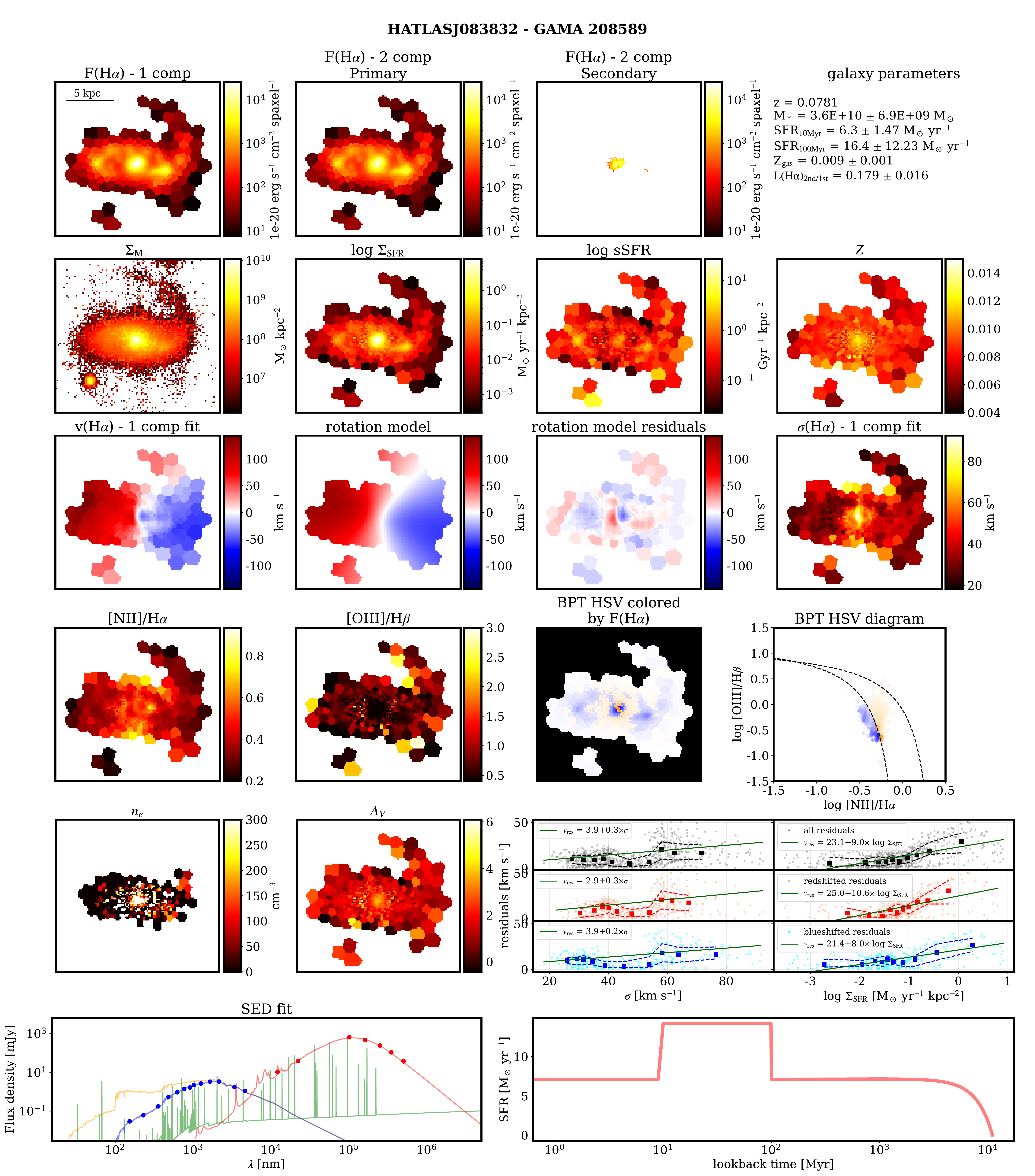}
\caption{Same as Fig.~\ref{fig:HATLASJ083601}, but for galaxy HATLASJ083832.}
\label{fig:HATLASJ083832}
\end{figure*}

\begin{figure*}
\centering
\includegraphics[width=\textwidth]{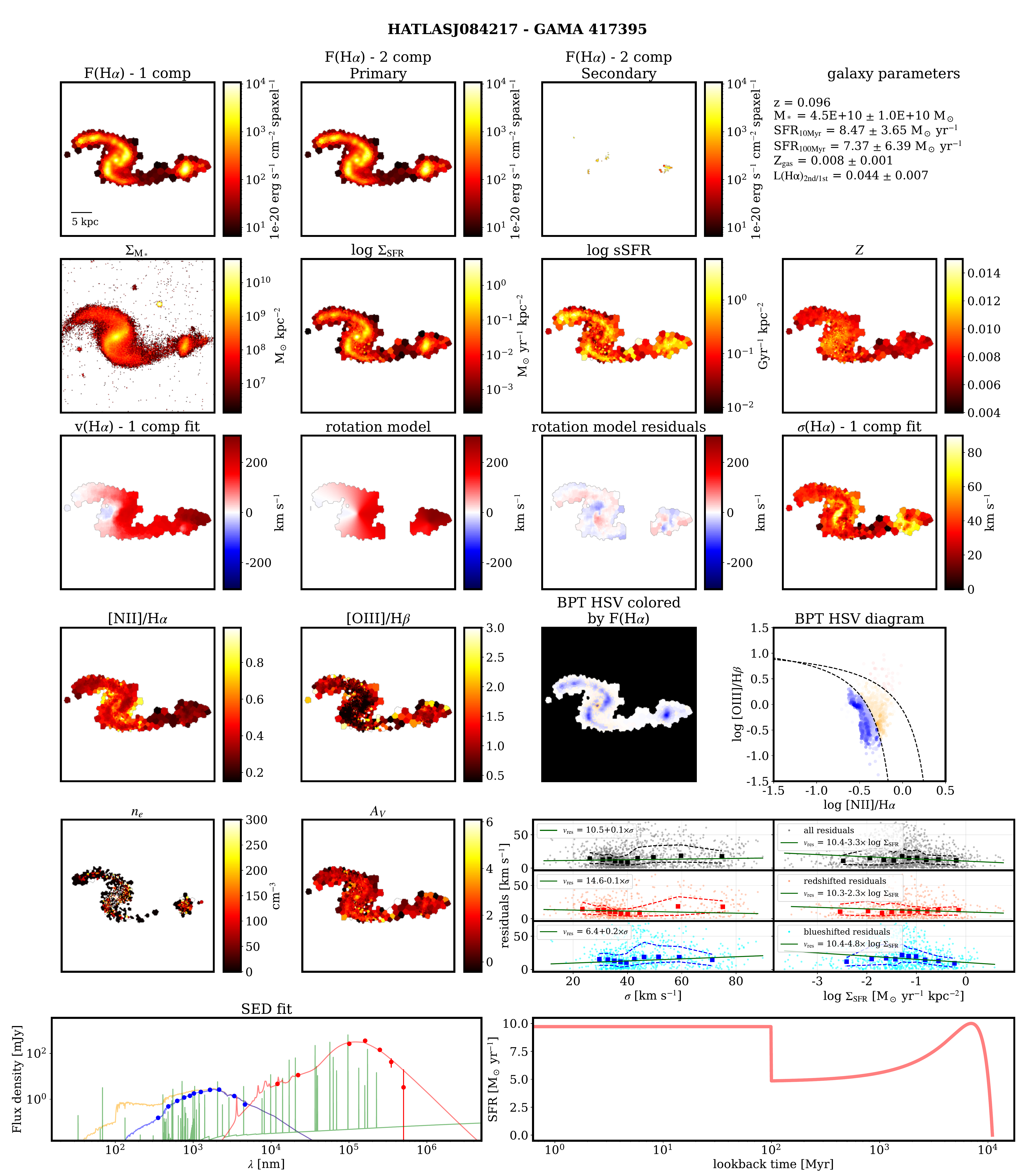}
\caption{Same as Fig.~\ref{fig:HATLASJ083601}, but for galaxy HATLASJ084217.}
\label{fig:HATLASJ084217}
\end{figure*}

\begin{figure*}
\centering
\includegraphics[width=\textwidth]{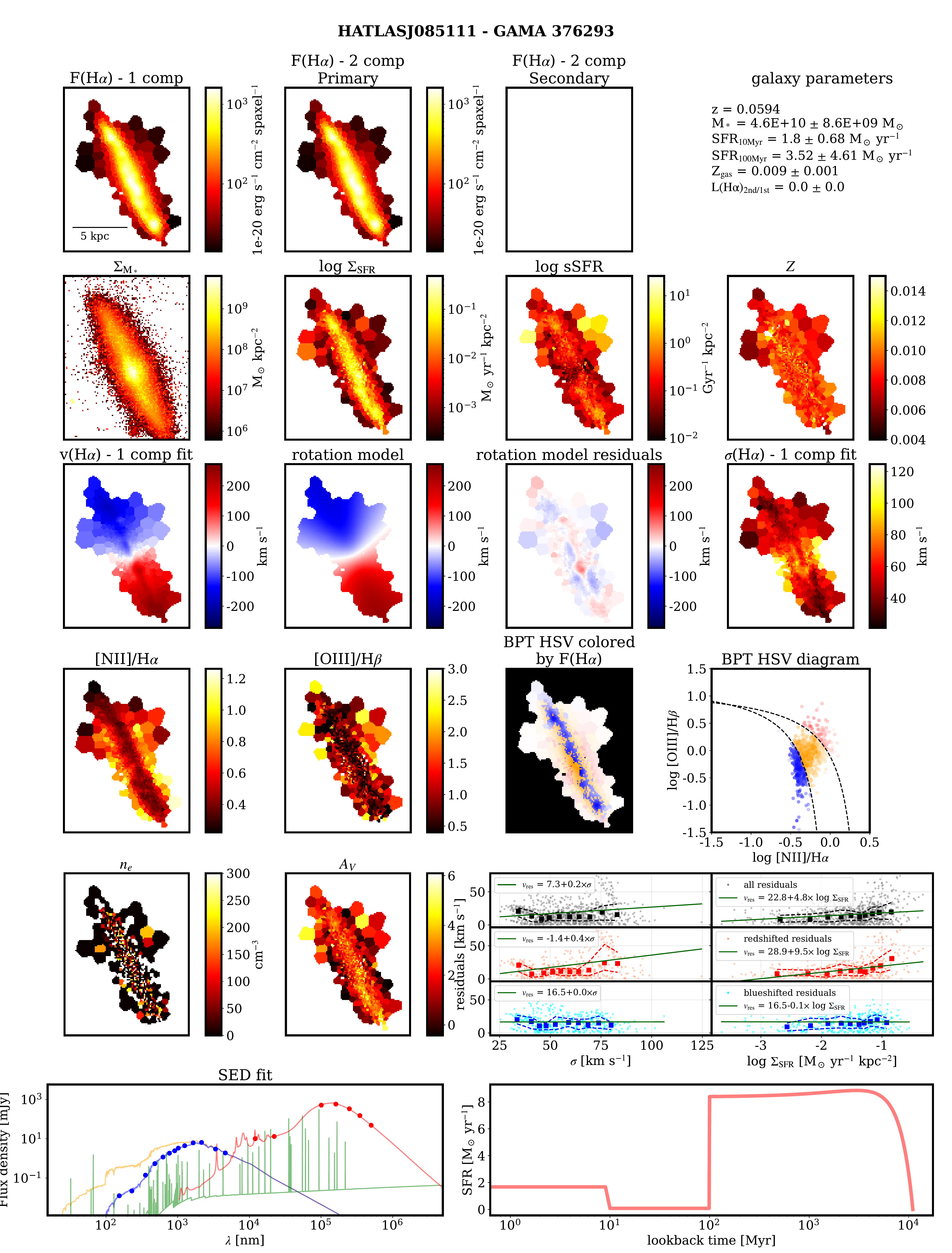}
\caption{Same as Fig.~\ref{fig:HATLASJ083601}, but for galaxy HATLASJ085111.}
\label{fig:HATLASJ085111}
\end{figure*}

\begin{figure*}
\centering
\includegraphics[width=\textwidth]{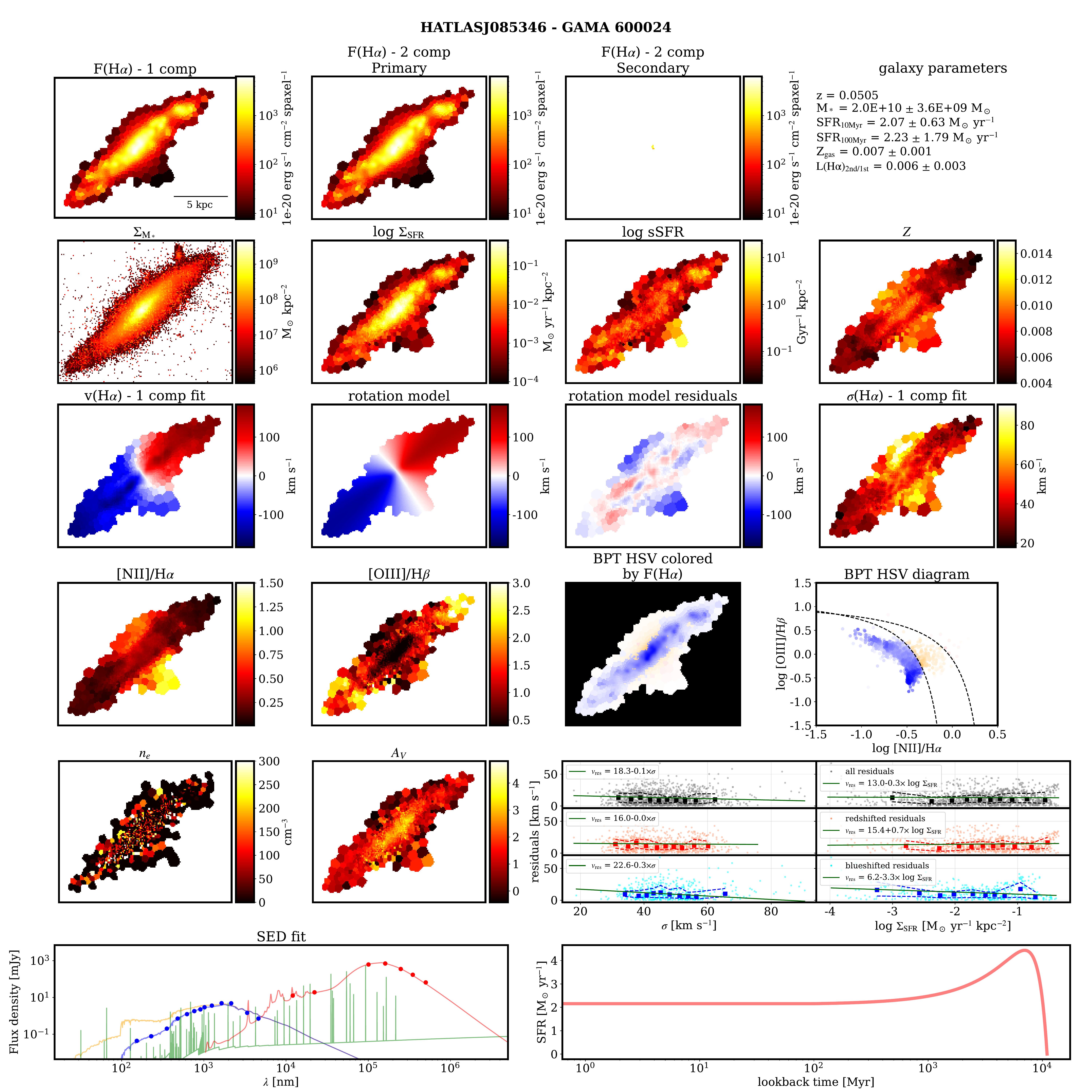}
\caption{Same as Fig.~\ref{fig:HATLASJ083601}, but for galaxy HATLASJ085346.}
\label{fig:HATLASJ085346}
\end{figure*}

\begin{figure*}
\centering
\includegraphics[width=\textwidth]{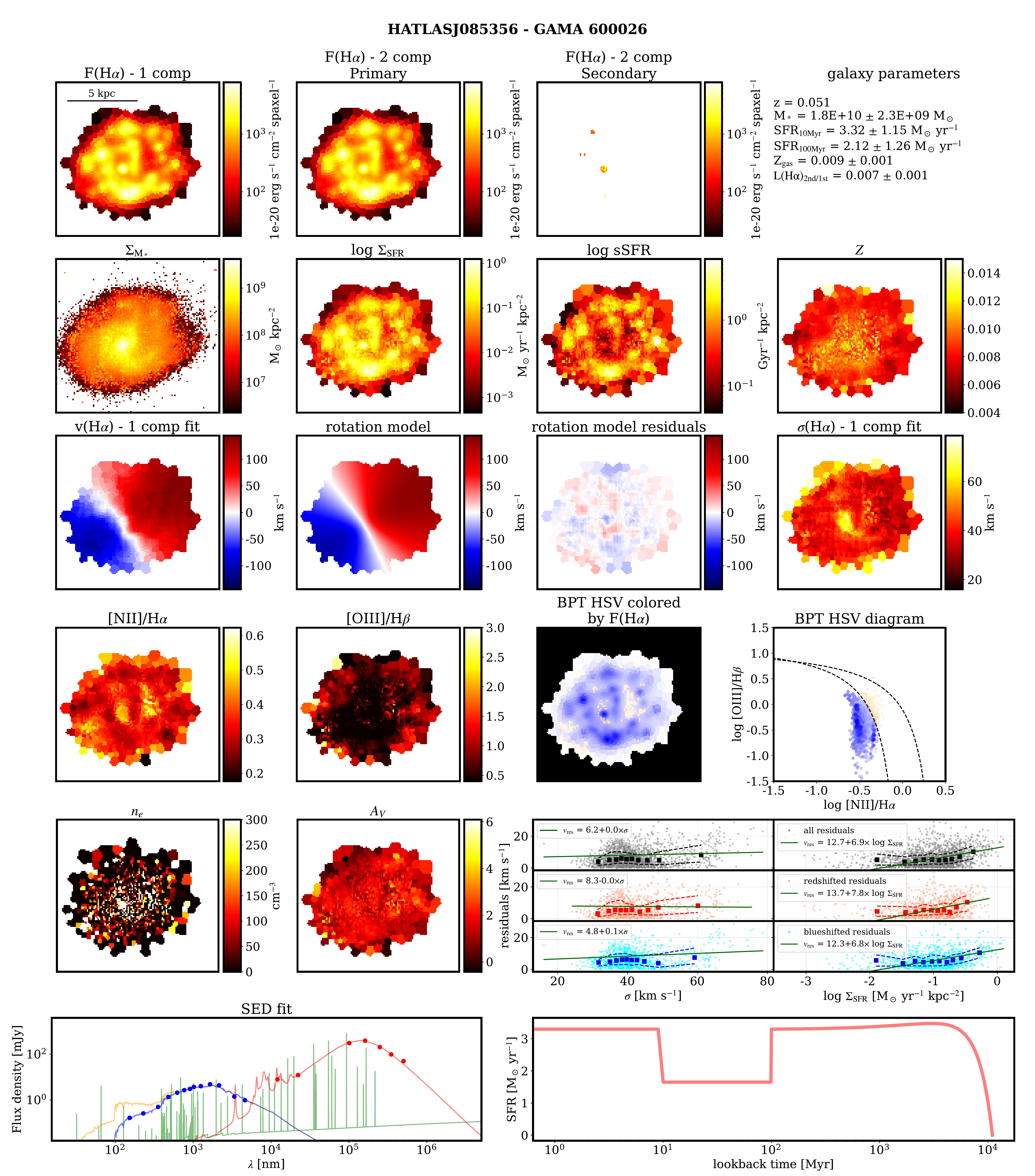}
\caption{Same as Fig.~\ref{fig:HATLASJ083601}, but for galaxy HATLASJ085356.}
\label{fig:HATLASJ085356}
\end{figure*}

\begin{figure*}
\centering
\includegraphics[width=\textwidth]{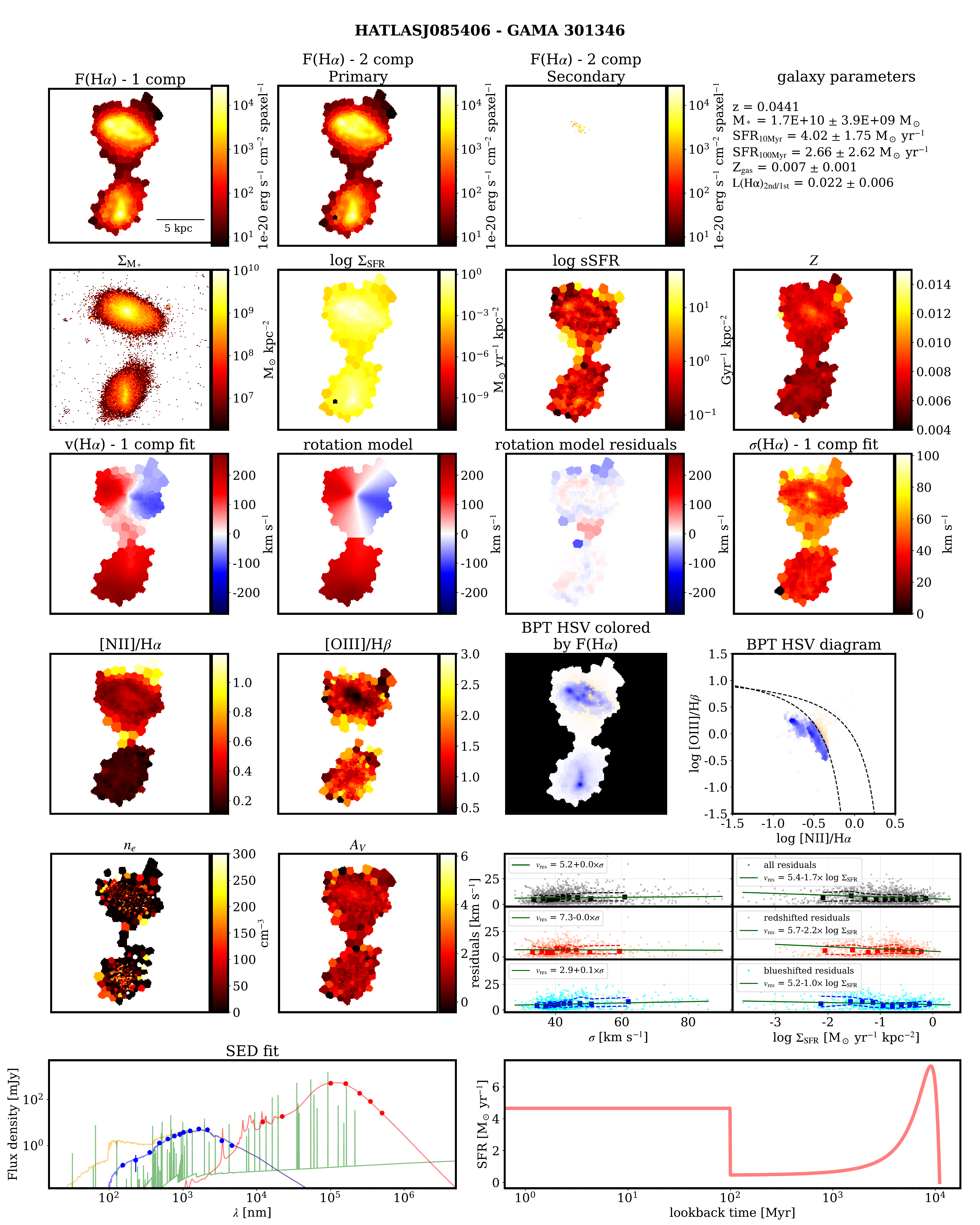}
\caption{Same as Fig.~\ref{fig:HATLASJ083601}, but for galaxy HATLASJ085406.}
\label{fig:HATLASJ085406}
\end{figure*}

\begin{figure*}
\centering
\includegraphics[width=\textwidth]{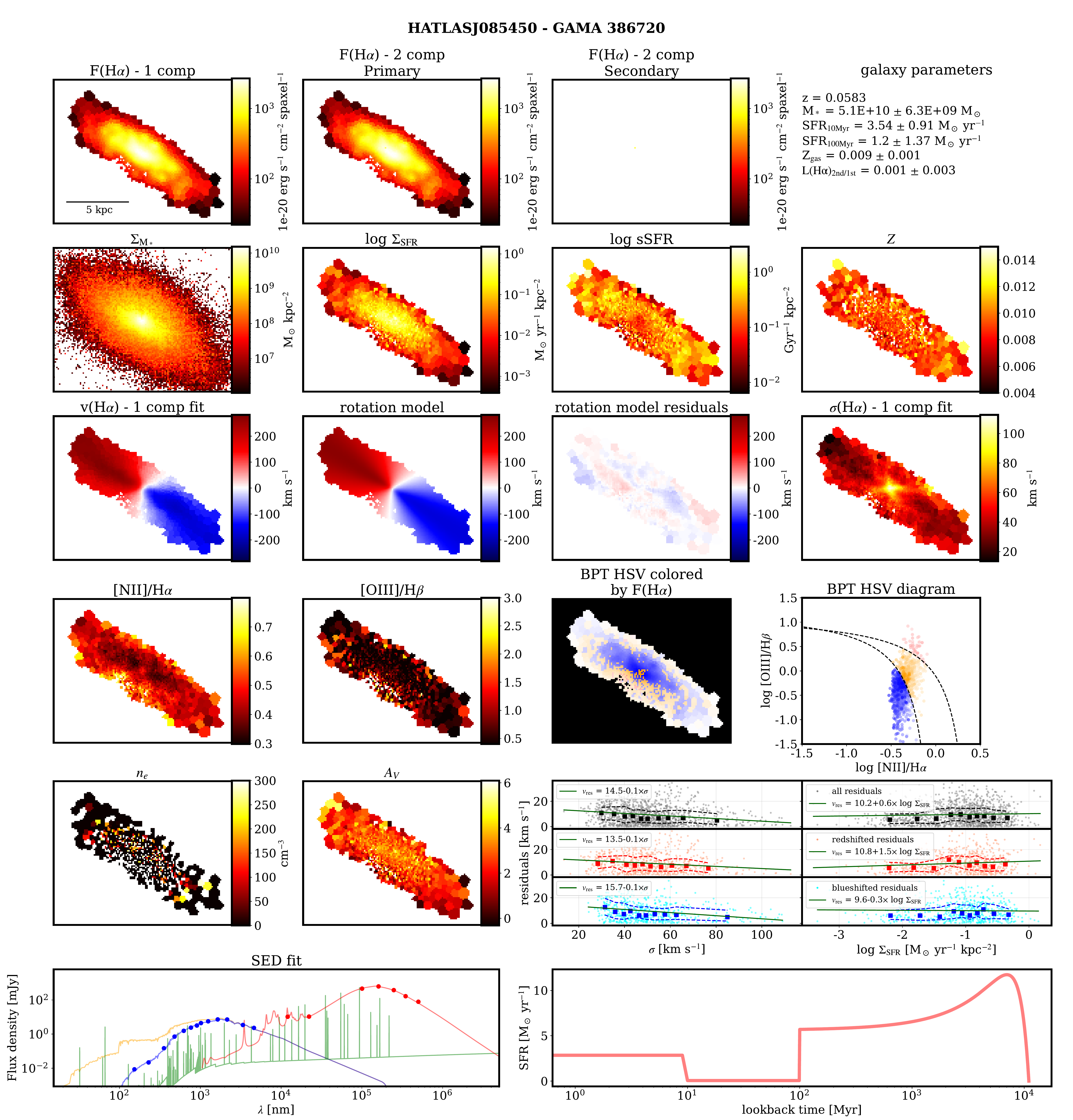}
\caption{Same as Fig.~\ref{fig:HATLASJ083601}, but for galaxy HATLASJ085450.}
\label{fig:HATLASJ085450}
\end{figure*}

\begin{figure*}
\centering
\includegraphics[width=\textwidth]{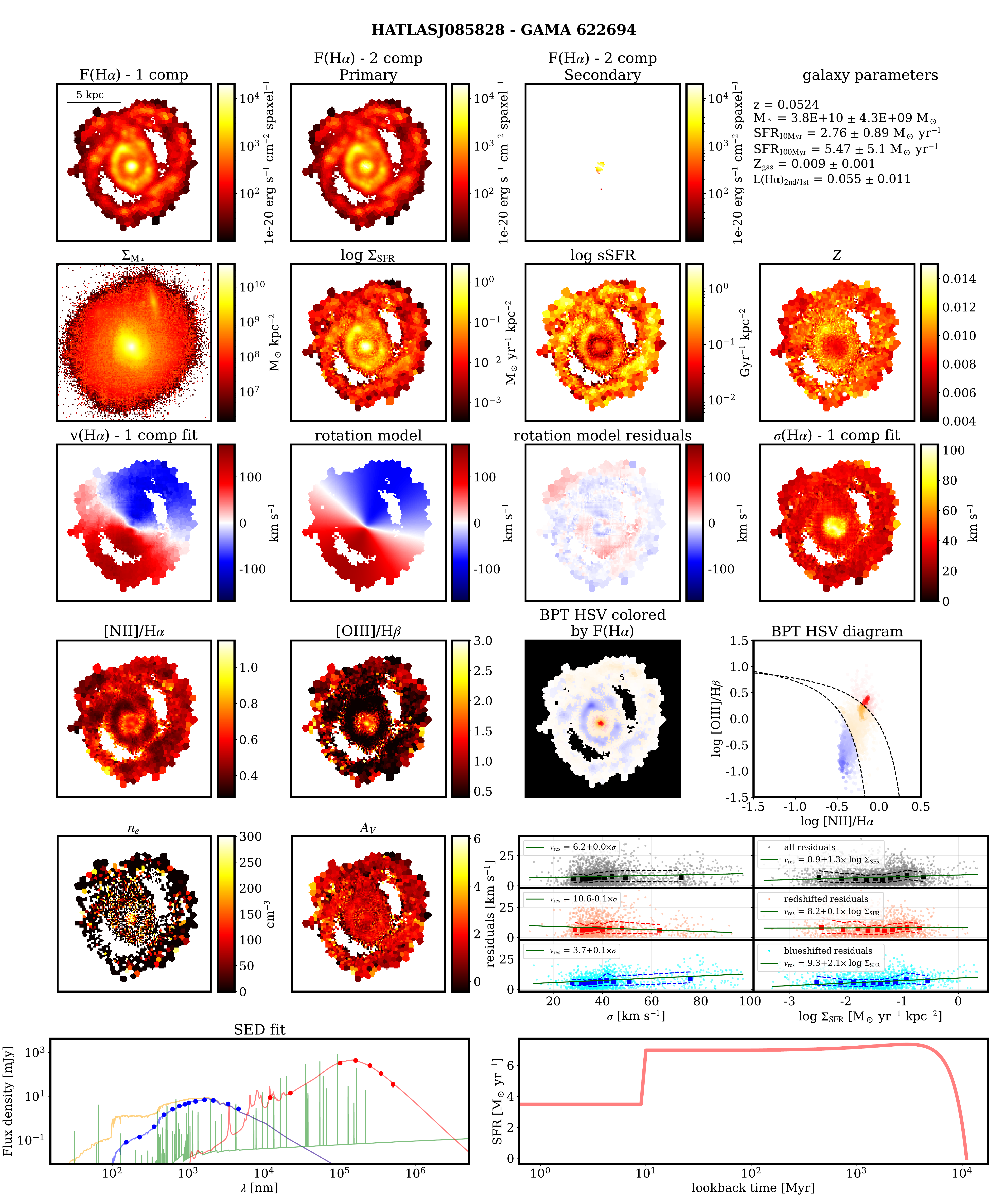}
\caption{Same as Fig.~\ref{fig:HATLASJ083601}, but for galaxy HATLASJ085828.}
\label{fig:HATLASJ085828}
\end{figure*}

\begin{figure*}
\centering
\includegraphics[width=\textwidth]{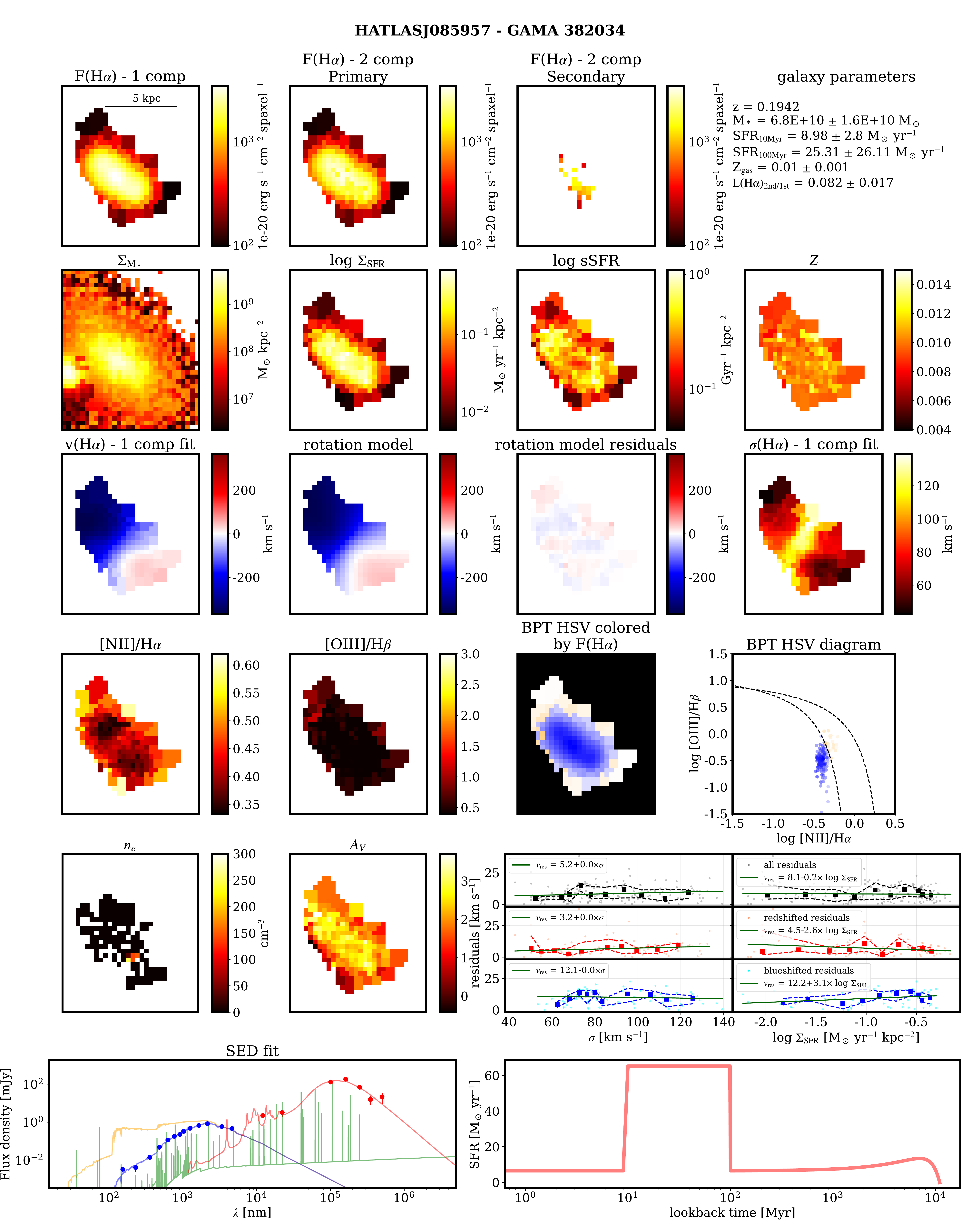}
\caption{Same as Fig.~\ref{fig:HATLASJ083601}, but for galaxy HATLASJ085957.}
\label{fig:HATLASJ085957}
\end{figure*}

\begin{figure*}
\centering
\includegraphics[width=\textwidth]{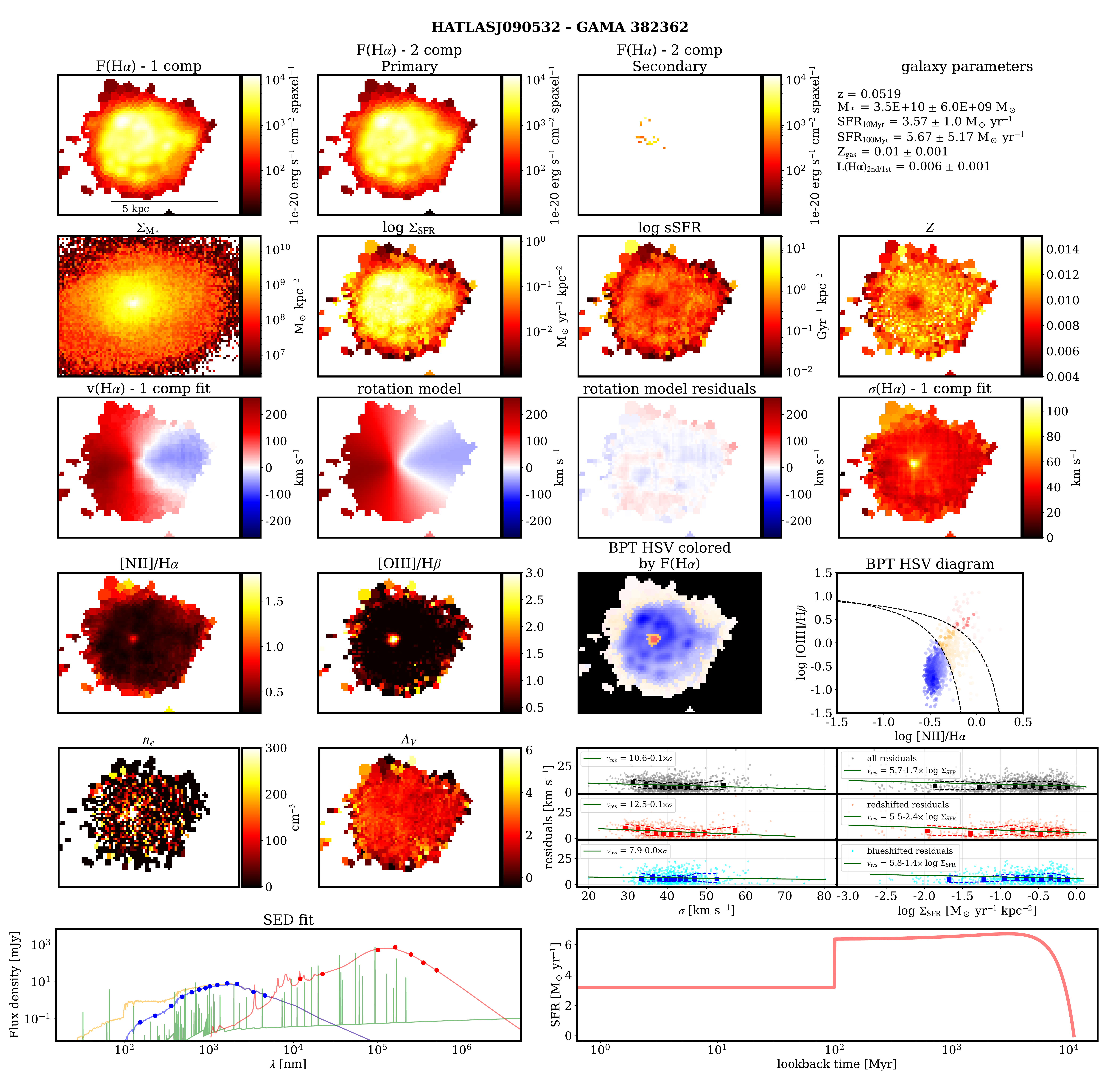}
\caption{Same as Fig.~\ref{fig:HATLASJ083601}, but for galaxy HATLASJ090532.}
\label{fig:HATLASJ090532}
\end{figure*}

\begin{figure*}
\centering
\includegraphics[width=\textwidth]{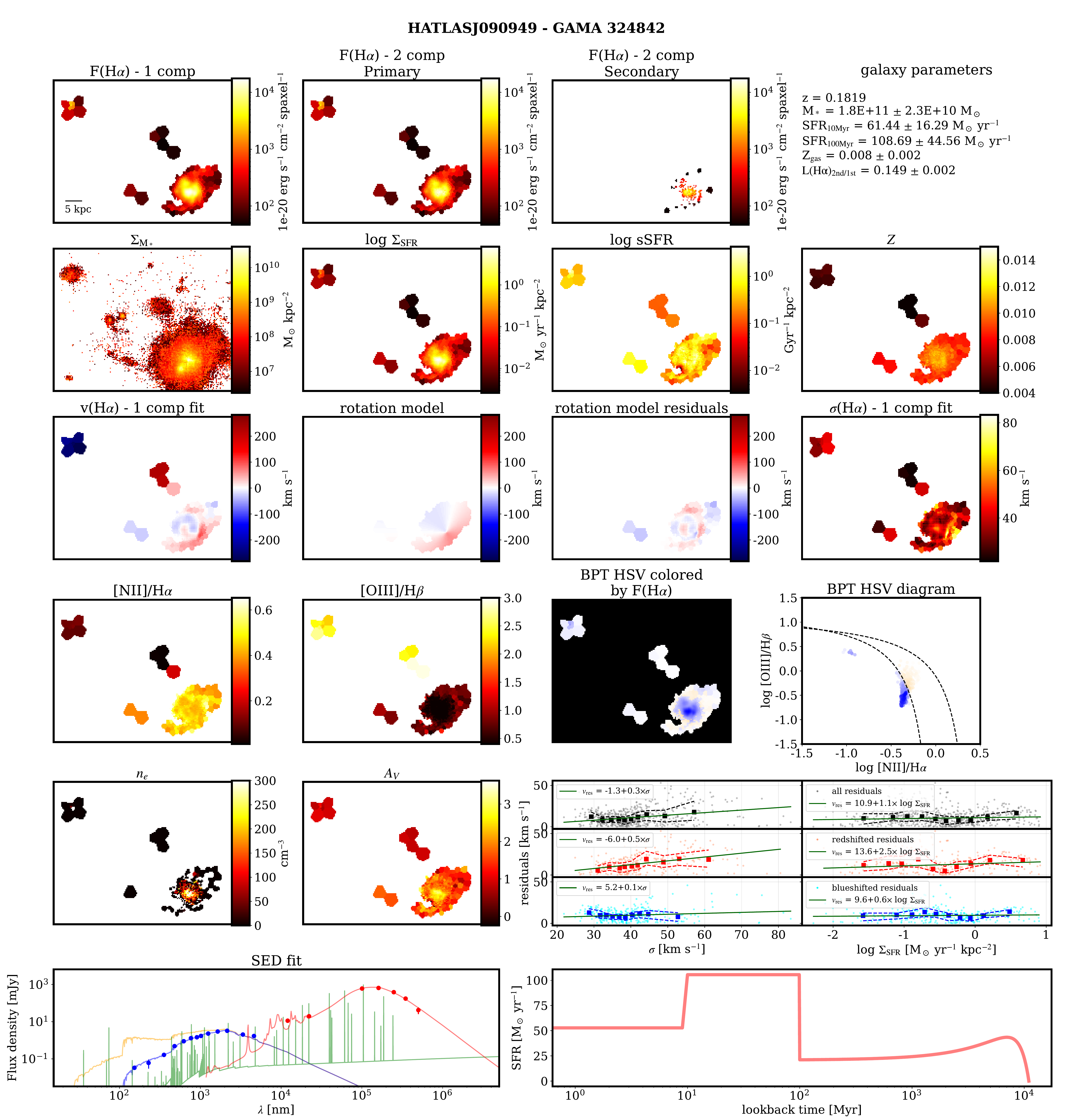}
\caption{Same as Fig.~\ref{fig:HATLASJ083601}, but for galaxy HATLASJ090949.}
\label{fig:HATLASJ090949}
\end{figure*}

\begin{figure*}
\centering
\includegraphics[width=\textwidth]{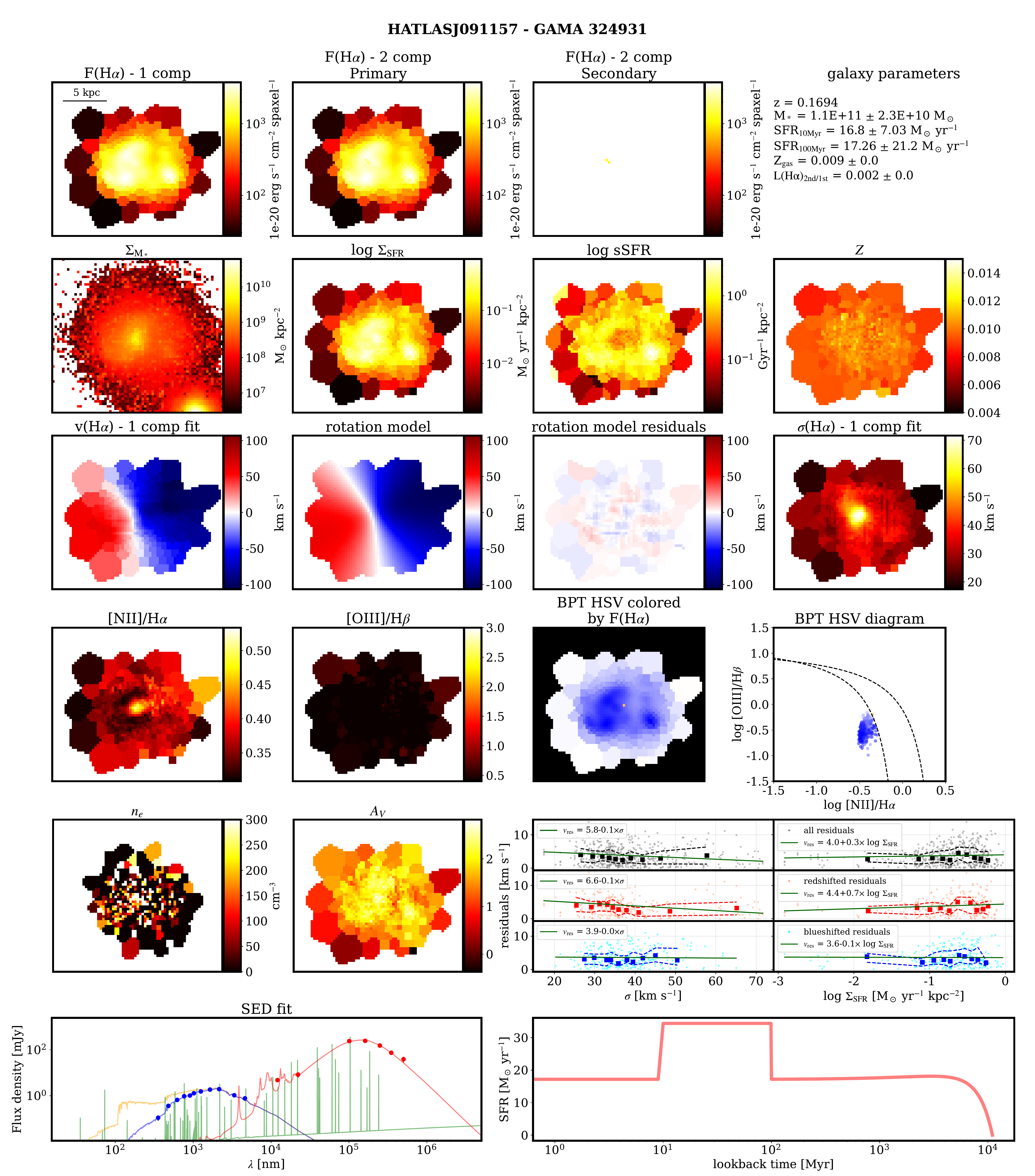}
\caption{Same as Fig.~\ref{fig:HATLASJ083601}, but for galaxy HATLASJ091157.}
\label{fig:HATLASJ091157}
\end{figure*}

\begin{figure*}
\centering
\includegraphics[width=\textwidth]{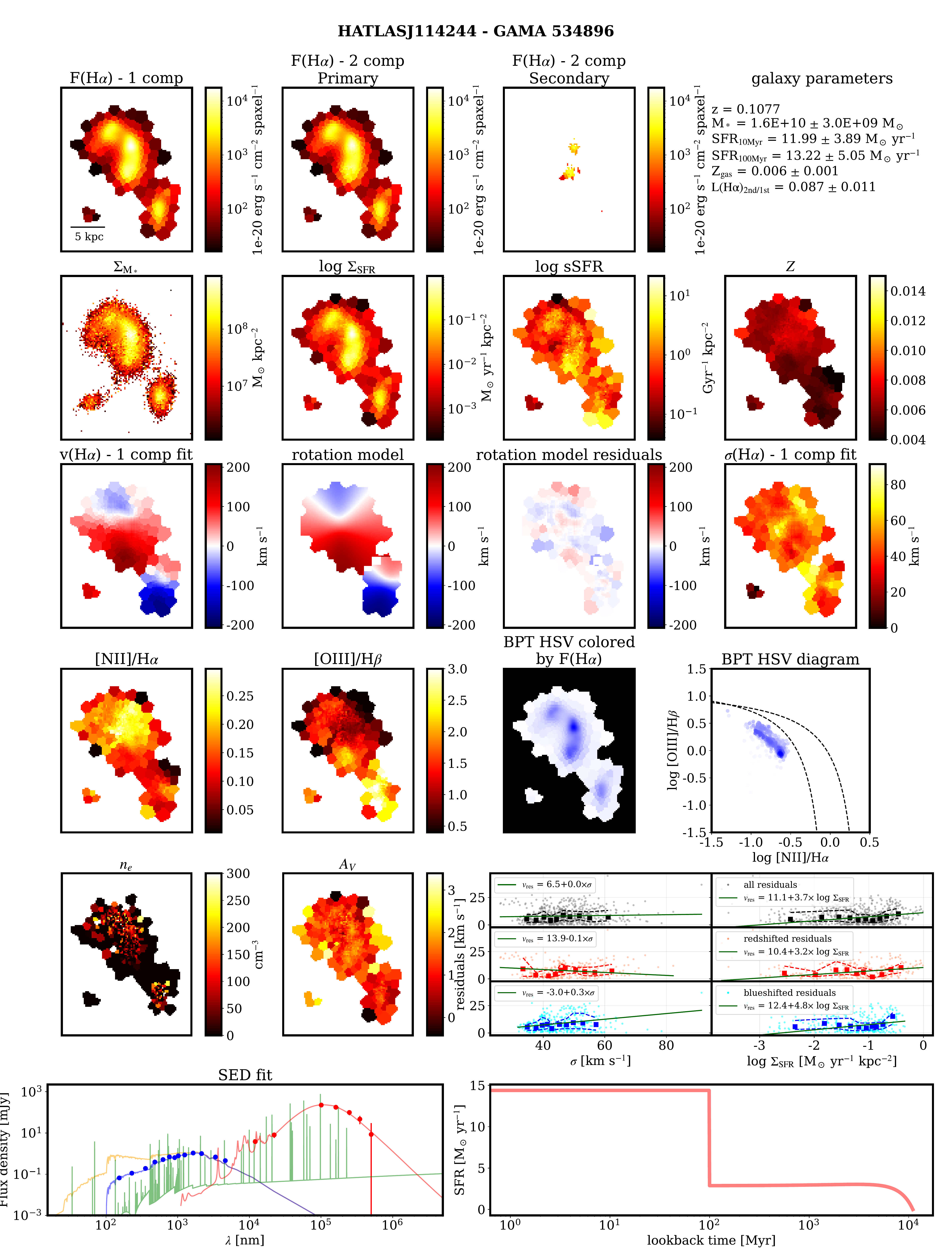}
\caption{Same as Fig.~\ref{fig:HATLASJ083601}, but for galaxy HATLASJ114244.}
\label{fig:HATLASJ114244}
\end{figure*}

\begin{figure*}
\centering
\includegraphics[width=\textwidth]{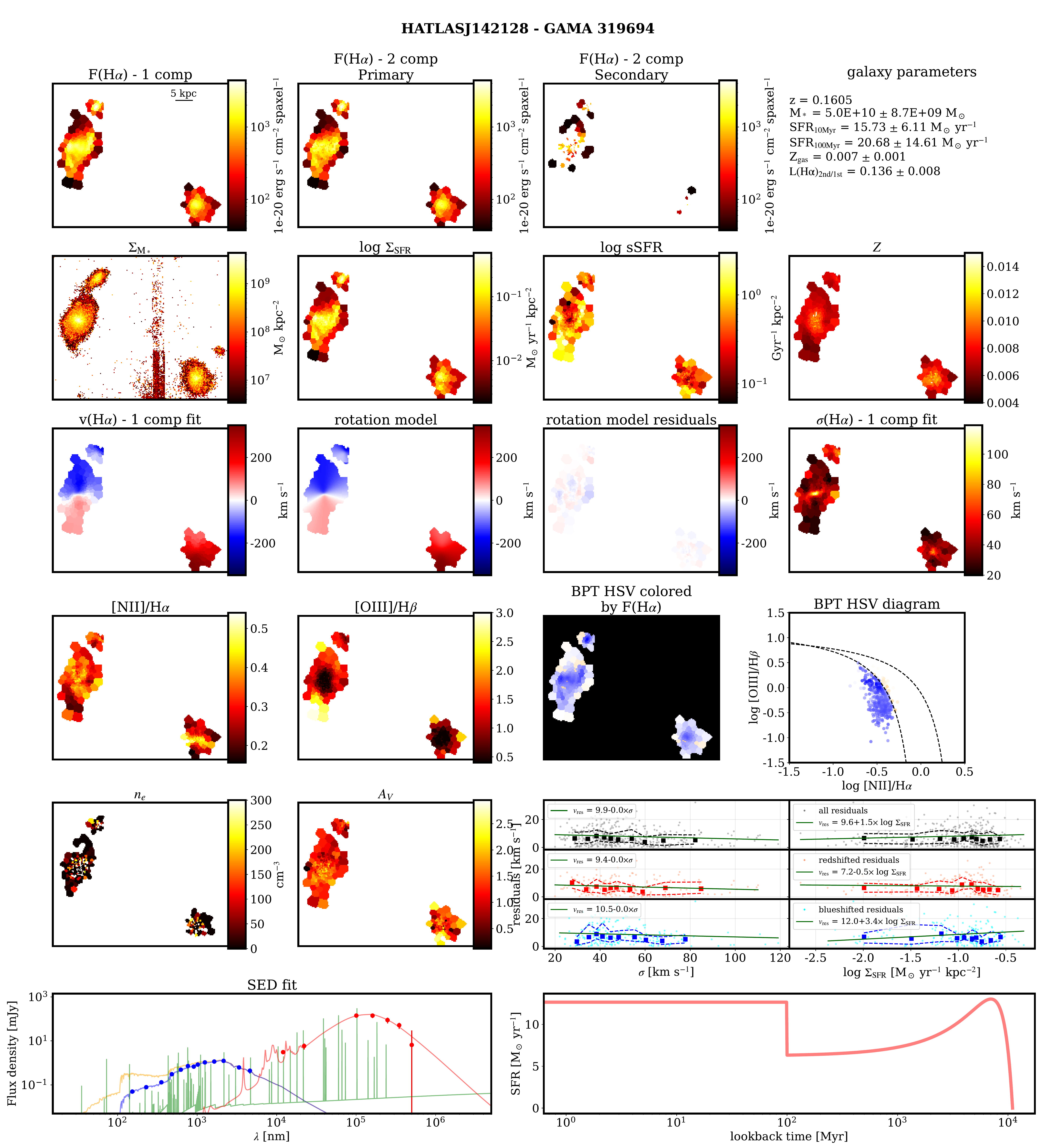}
\caption{Same as Fig.~\ref{fig:HATLASJ083601}, but for galaxy HATLASJ142128.}
\label{fig:HATLASJ142128}
\end{figure*}

Here we present relevant information for each individual galaxy, shown in Figs. \ref{fig:HATLASJ083601}-\ref{fig:HATLASJ142128}, including distribution maps, physical parameters and SED and BPT plots. The caption in Fig.~\ref{fig:HATLASJ083601} details these information and applies to the other figures.

\end{document}